\documentclass[structabstract]{aa}  

\usepackage{graphicx}
\usepackage{epstopdf}
\usepackage{color}

\usepackage{txfonts}
\usepackage{natbib}
\bibpunct{(}{)}{;}{a}{}{,} 
\def\gr{$\gamma$-ray}

\begin{document}
\title{EGMF Constraints from Simultaneous GeV-TeV Observations of Blazars}
 
\author{A.~M.~Taylor 
          \inst{1},
          I.~Vovk
          \inst{1}
          \and
          A.~Neronov
          \inst{1}}

\institute{ISDC Data Centre for Astrophysics, Ch. d'Ecogia 16, 1290, Versoix, Switzerland \\
\email{Andrew.Taylor@unige.ch}}

\abstract
   {Attenuation of the TeV \gr\ flux from distant blazars through pair production with extragalactic 
background light leads to the development of electromagnetic cascades and subsequent, lower energy, 
GeV secondary \gr\ emission. 
Due to the deflection of VHE cascade electrons by extragalactic magnetic fields (EGMF), the spectral 
shape of this arriving cascade \gr\ emission is dependent on the strength of the EGMF. Thus, 
the spectral shape of the GeV-TeV emission from blazars has the potential to probe the EGMF strength
along the line of sight to the object. Constraints on EGMF previously derived from the gamma-ray data suffer from an uncertainty related to the non-simultaneity of GeV and TeV band observations. }
   {We investigate constraints on the EGMF derived from observations of blazars for which TeV observations simultaneous with  those by \textit{Fermi} telescope were reported. We study the dependence of the EGMF bound on the hidden assumptions
it rests upon.}
   {We select blazar objects for which simultaneous Fermi/LAT GeV and Veritas, MAGIC or HESS TeV emission
have been published. We model the development of electromagnetic cascades along the gamma-ray beams from these sources using  Monte Carlo simulations, including the calculation of  the temporal delay
incurred by cascade photons, relative to the light propagation time of direct $\gamma$-rays from the source. }
   {Constraints on EGMF could be derived from the simultaneous GeV-TeV data on the blazars RGB~J0710+591, 1ES~0229+200, and 1ES~1218+304.  The measured source flux level in the GeV band is lower than the flux of the expected cascade component calculated under the assumption of zero EGMF. Assuming that the reason for the suppression of the cascade component is the extended nature of the cascade emission, we find that $B\gtrsim 10^{-15}$~G (assuming EGMF correlation length of $\ge 1$ Mpc) is consistent with the data. Alternatively, the assumption that the suppression of the cascade emission is caused by the time delay of the cascade photons the data are consistent with  $B\gtrsim 10^{-17}$~G for the same correlation length.  }
   {}


\maketitle

\section{Introduction}
\label{Introduction}

The presence of magnetic fields in galaxies and galaxy clusters  plays a key role in present day astrophysical studies.  However, the origin of these fields remains largely uncertain (see \cite{kronberg94,grasso00,widrow02,beck08} for reviews). A commonly accepted hypothesis is that relatively strong galactic and cluster magnetic fields  result from the amplification of much weaker pre-existing  ``seed'' fields via compression and turbulence/dynamo amplification in the course of structure formation processes \citep{kulsrud}. 

The origin of these seed magnetic fields is unknown. It is possible that the seed fields are produced locally in (proto)galaxies via the so-called ``Biermann battery'' mechanism \citep{pudritz89,gnedin00}. Otherwise, the seed fields might be of primordial origin, i.e. produced at the moments of phase transitions in the Early Universe \citep{grasso00,widrow02}. Constraints on the nature of the seed fields could potentially be derived from the measurements of weak magnetic fields in the intergalactic medium which are not amplified by the action of different types of dynamos.

The measurement of extremely weak magnetic fields in the voids of the Large Scale Structure (LSS) is a challenging task and up to now only upper  bounds have been derived using various techniques. The tightest upper bounds come from the  search for the Faraday rotation of polarization of radio emission from distant quasars \citep{kronberg76,kronberg_perry,blasi} and from the effect of magnetic fields on the anisotropy of Cosmic Microwave Background radiation \citep{barrow,durrer00}. 

A new  handle on the EGMF measure, using the cascade emission from blazars, is now emerging as an alternative probe. In this method, multi-TeV $\gamma$-rays from distant ($>100$~Mpc) blazars attenuate through pair production interactions on the extragalactic background light (EBL), leading to the development of electromagnetic cascades  \citep{aharonian94,Plaga:1995,Coppi:1996ze,neronov07,d'Avezac:2007sg,japanese,Eungwanichayapant:2009bi,neronov09,elyiv09,kachelriess09}. The angular pattern of the secondary cascade emission from $e^+e^-$ pairs deposited in the intergalactic medium through pair production interactions depends on the EGMF strength. The detection (non-detection) of the cascade emission signal from known TeV $\gamma$-ray emitting blazars could result in the measurement of (lower bound on) the strength of the magnetic field in intergalactic space along the line of sight toward these blazars. The first application of this method for  deriving lower bounds on the EGMF have been carried out  \citep{Neronov:2009,Tavecchio:2010ja,Dolag:2010ni,Dermer:2010mm}, suggesting that a measure of the EGMF may finally soon be within reach. 

In the simplest settings, the lower bounds on EGMF at the level of $10^{-17}$ to $10^{-15}$~G \citep{Neronov:2009,Tavecchio:2010ja,Dolag:2010ni} (depending on assumptions about intrinsic blazar spectra) were derived. These bounds adopt a simplifying assumption that the measured blazar fluxes provide correct estimates of the time-averaged fluxes in the GeV and TeV energy bands. This assumption could, in principle,  be partially verified via a systematic monitoring of the sources simultaneously in the TeV (for primary source emission) and GeV (for the cascade emission) energy bands. 

In this work, we consider a set of blazars observed simultaneously both by Fermi/LAT and HESS or Veritas, in order to search for the cascade component of the GeV-TeV spectra of these sources. We find that in several cases, namely for the blazars RGB~J0710+591 and 1ES~1218+304, the measured source flux in the GeV band is lower than the expected (minimal possible) flux produced by the gamma-ray cascade in intergalactic space, calculated assuming zero magnetic field along the line of sight. This imposes a lower bound on the strength of magnetic field in the intergalactic medium, similar to the bounds found for the cases of 1ES~0229+200 \citep{Neronov:2009,Tavecchio:2010ja,Dolag:2010ni} and 1ES~0347-121 \citep{Neronov:2009}. However, contrary to the bounds from 1ES~0229+200 and 1ES~0347-121, no additional assumptions about long-term stability of the source in the GeV and TeV bands are needed,
because of the truly simultaneous nature of multi-band observations\footnote{Note that because of the difference in observation techniques, GeV and TeV measurements are accumulated at different time scales (year for the GeV band data and night-by-night in the TeV band).}. 

Availability of the simultaneous data allows a study of two alternative possible reasons for the suppression of the cascade signal: dilution of the cascade flux due to the time delay of the cascade signal, following a period of enhanced activity of the source vs. suppression of the cascade signal contribution to the point source flux due to the extended nature of the cascade emission.  We show that adopting an (possibly extreme) assumption that the source gamma-ray activity periods of the sources are as short as $\sim 2$~yr (the time scale of {\it Fermi} operation), the lower bound on the magnetic field relaxes to  $\sim 10^{-18}$~G for the cases of RGB~J0710+591 and 1ES~1218+304. 

When this paper was almost ready for publication a report on investigation of constraints on EGMF derived assuming suppression of the cascade emission due to the time delay of the cascade flux appeared \citep{Dermer:2010mm}. The analysis of \citet{Dermer:2010mm} relies  on an unpublished VERITAS observation of 1ES~0229+200 \citep{perkins10} which  reveals that the flux and spectrum of the source did not change on a 4-year time span between HESS \citep{0229_HESS} and VERITAS \citep{perkins10} observations. Relying on this statement, we include 1ES~0229+200 in our analysis, to verify the results obtained by \citet{Dermer:2010mm}. We find that simplifying assumptions adopted in the analytical  modeling of electromagnetic cascade by \citet{Dermer:2010mm} led to a large underestimate of the lower bound on EGMF. Correction of the result of \citet{Dermer:2010mm} found using full Monte-Carlo simulation of the electromagnetic cascade  we find a lower bound $B\ge 10^{-17}$~G from the minimal possible time delay of the cascade signal in 1ES~0229+200.

\section{Selection of sources and data analysis}
\label{Selection}

Detailed calculation of the spectrum of GeV-band emission from electromagnetic cascade initiated by absorption of very-high-energy $\gamma$-rays on the EBL requires the knowledge of the initial (unabsorbed) source flux in the TeV energy band.  If the magnetic field in the intergalactic medium is close to zero, the cascade GeV $\gamma$-rays arrive almost simultaneously with the primary absorbed TeV $\gamma$-rays, with only a small ($\sim 10$~hr scale) magnetic field-independent time delay related to the angular scatter of electrons and positrons in the pair production process \citep{neronov09}. This means that prediction of the level of the GeV-band cascade emission at a given moment of time requires the knowledge of the simultaneous TeV flux of the source. 

The Large Area Telescope (LAT) on board of {\it Fermi} satellite has performed continuous monitoring of the entire sky in the GeV band since August 4, 2008. This implies that any blazar observations in the TeV band, performed with the ground-based gamma-ray telescopes HESS, MAGIC or Veritas after August 4, 2008 automatically have simultaneous observations in the GeV band. Several  blazar observations performed after August 4, 2008 were reported: Mrk~501 \citep{Mrk_501_Veritas}, PKS~2155-304 \citep{PKS_2155_HESS}, 3C~66A at \citep{3C66A_Veritas}, RGB~J0710+591 \citep{RGB_J0710_Veritas} and 1ES~1218+304 \citep{1ES_1218_Veritas}, 1ES~0229+200 \citep{Dermer:2010mm,perkins10} and PKS 1424+240 \citep{pks1424_Veritas}. Our analysis of constraints on EGMF is based on the analysis of  GeV-TeV band  spectral properties of the sources listed above.
 
For each of the sources with simultaneous GeV-TeV data, we derive the spectral characteristics in the 0.1-100~GeV band from {\it Fermi}/LAT data and combine them with the reported TeV band spectrum  to produce the broad band spectra.  In our analysis of {\it Fermi}/LAT data we use the publicly available data from the Fermi/LAT from August 4, 2008 till November 30, 2010. We process the data using {\it Fermi} Science Tools package 
of the version v9r17p0\footnote{http://http://fermi.gsfc.nasa.gov/ssc/data/analysis/}. For the spectral extraction we use the unbinned likelihood analysis method, taking into account all the sources from the first year Fermi catalog \citep{1st_year_catalog} situated within an angle $\le 10^\circ$ from the blazar of interest.  We model the broad band (0.1~GeV -10~TeV) spectra of the sources with a two-component model containing intrinsic emission from the primary source and emission from an electromagnetic cascade initiated by the absorption of very-high-energy gamma-ray interactions with the EBL, as explained in the next section.

\begin{table}[htcb]
 \centering
 \begin{tabular}{|l|l|l|l|l|l|}
   \hline
   Name          & $RA$ & $DEC$ & $z$ & $\Gamma$ & $E_{\rm cut}$ \\
   \hline
   1ES~0229+200  & $38.203^\circ$  & $20.288^{\circ}$ & 0.14 &1.2 &5.0 \\
   RGB~J0710+591 & $107.625^\circ$  & $59.139^\circ$ & 0.13 &1.6 &1.0 \\
   1ES~1218+304  & $185.341^\circ$ & $30.177^{\circ}$ & 0.18 &1.7 &2.5 \\
   \hline
 \end{tabular}
 \caption{Blazars considered in the analysis leading to the lower bound on EGMF. $\Gamma$ and $E_{\rm cut}$ are the  limiting values of  photon index and the cut-off energy derived from the fit of the Fermi and TeV data with the direct emission (cut-off powerlaw attenuated by the pair production on the EBL) plus cascade model, under the requirement of a minimal cascade contribution. The cut-off energy in in units of TeV.}
 \label{sources_list}
\end{table}

For PKS~1424+240 the uncertainty of the source redshift does not allow firm predictions of the amount of cascade emission to be made. Taking this into account, we exclude this source from our analysis. For three sources, Mrk~501, 3C~66A, and PKS~2155-304, the slopes of the intrinsic spectra found in the fitting procedure are relatively soft, with photon indices  $\Gamma\simeq 2$. In such settings, the cascade emission gives a sub-dominant contribution to the source spectrum in the GeV band and no sensible constraints on the magnetic field along the line of sight could be derived from the analysis. The remaining three sources, listed in Table \ref{sources_list}, have harder intrinsic spectra. 
We note that quite a small range of blazar redshifts is left, as seen in Table 1, resulting from the 
application of the selection criterias, mentioned above. For a given redshift bin size, there is more 
chance of finding a hard-spectrum blazar for the larger redshift values. On the other hand, for distant 
blazars absorption on the EBL in very significant and may cause them to become invisible in the TeV 
band. The combination of the above mentioned effects leads to the selection of the set of three sources,
listed in Table 1 with quite a narrow redshift range.
The analysis of the data on these three sources constrains the EGMF. 
\section{Monte Carlo simulations}
\label{Monte_Carlo}

Extragalactic electromagnetic cascades, in the presence of
non-negligible EGMF ($>10^{-20}$~G), evolve both spatially and
energetically as the propagation front of the emission moves
away from the source. The nature of the constraints on the EGMF derived from the timing and imaging analysis of the signal produced by electromagnetic cascades developing in intergalactic space can be qualitatively understood from the decomposition of the cascade signal in space and time, which we illustrate using Monte-Carlo simulations.

In order to convey the key features  introduced by the spatial evolution of  the cascade, we start with a 
consideration of simplified situation of a collimated primary gamma-ray beam (i.e. neglecting the finite jet opening angle). To further simplify the consideration, we consider an idealized situation  in which a distant source injects primary gamma-rays at a fixed energy $E_0=100$~TeV. The energy $E_0$ is chosen in such a way that the optical depth with respect to pair production on the EBL is $\tau(E_0)\gg 1$. The resulting spectra of the cascade emission depend only weakly on $E_0$ as soon at $\tau(E_0)\gg 1$. In these illustrative calculations we assume the source redshift $z=0.13$, equal to the redshift of RGB~J0710+591. Later in this section we introduce non-zero intrinsic jet opening angle $\theta_{jet}>0$ and broad band emission spectrum and  describe how our simplified results are altered. 

\begin{figure}
\begin{center}
\includegraphics[width=0.518\columnwidth,angle=-90]{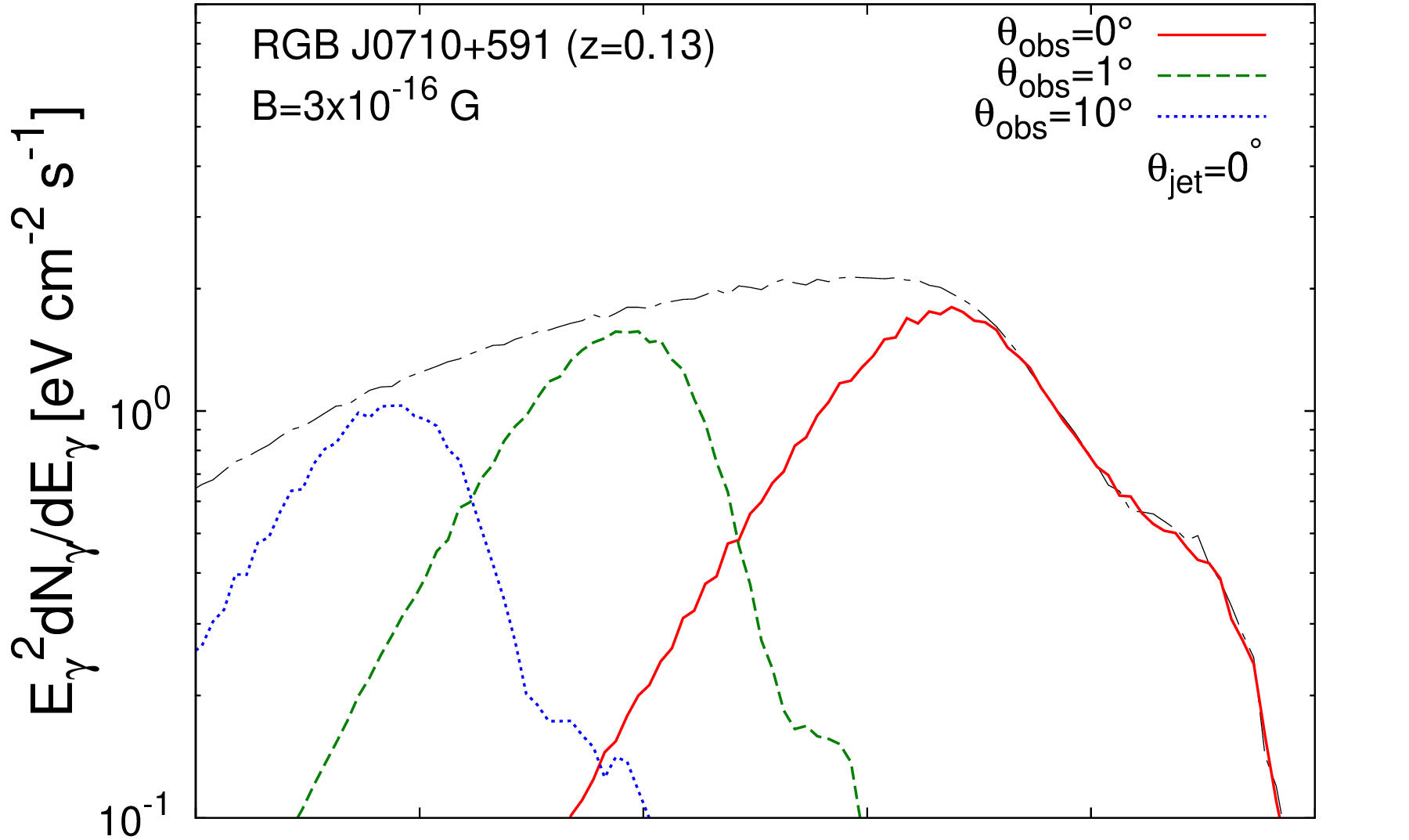}
\includegraphics[width=0.6\columnwidth,angle=-90]{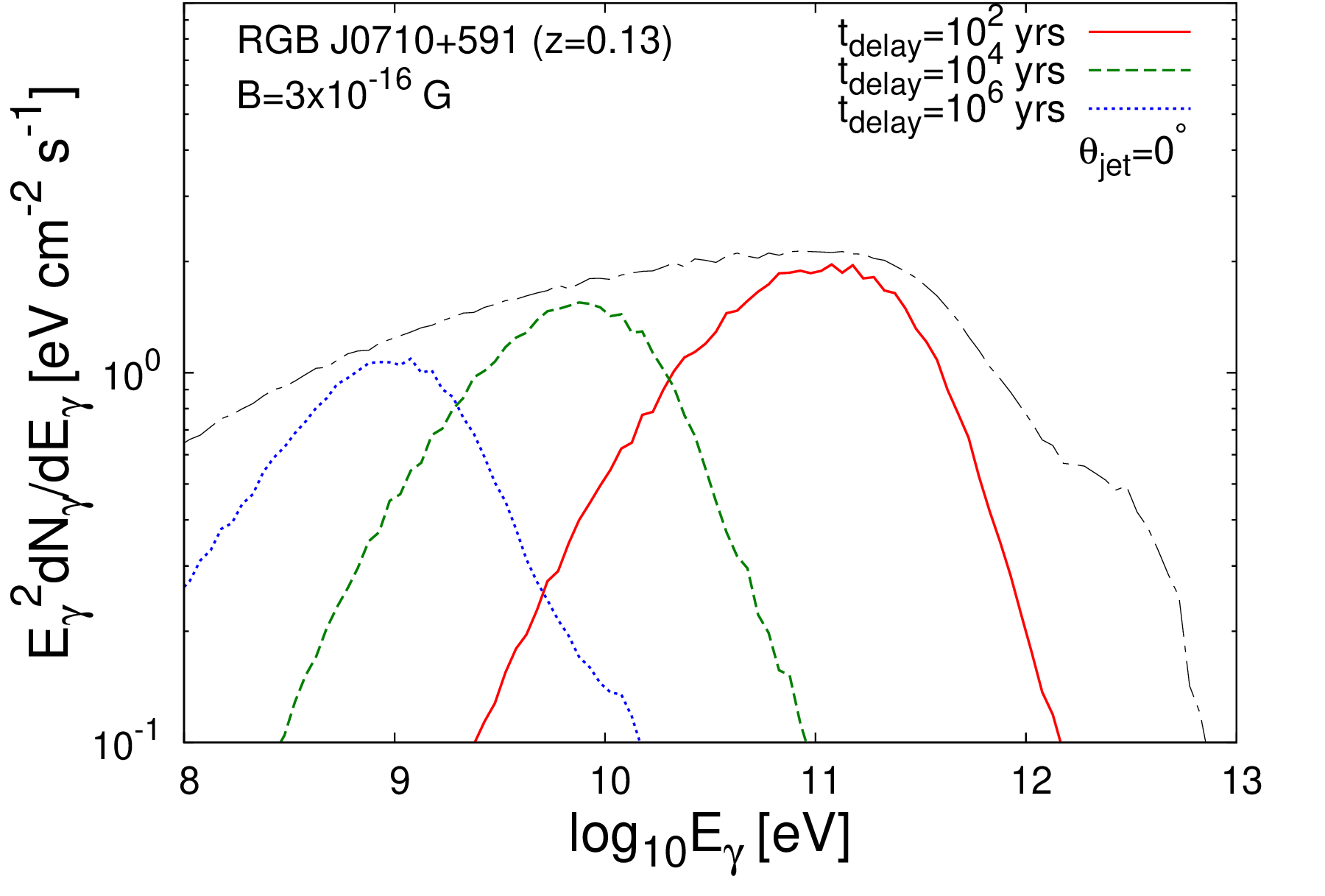}
\caption{The arriving energy fluxes following the injection of a 10$^{14}$~eV photon flux from a source at redshift $z=0.13$, with an intervening EGMF=$3\times 10^{-16}$G. Top: The arriving flux is decomposed into that observed from different observer angles, $\theta_{\rm obs}$. Bottom: The arriving flux is decomposed into that arriving with different delay times, $t_{\rm delay}$. In both plots, the long-short dash line represents the envelope flux containing the non-decomposed spectrum.}
\label{angle_time_cuts}
\end{center}
\end{figure}

Fig. 1 shows the results of the model calculation.  The arriving spectra of the cascade emission, for the case in which the cascade
develops in the presence of a negligible EGMF,
is the long dash, short dash line 
seen in both panels of figure~\ref{angle_time_cuts}.
This arriving photon spectrum may, in fact, be obtained simply using a 
``kinetic equation'' description. Indeed, we have compared our results using a 
Monte Carlo description used in this work (for more details see 
\cite{Taylor:2008jz}) and the kinetic description 
results for the case of a negligible B-field, used in \citet{Neronov:2009}, finding good
agreement in all cases. 

For these results and the results throughout this paper, the EBL model of \citet{Franceschini:2008tp} is adopted, in 
which the evolution of the EBL with redshift is accounted for. Furthermore, 
the cosmological parameters of $H_{0}=70$~km~s$^{-1}$~Mpc$^{-1}$, 
$\Omega_{M}=0.3$, and $\Omega_{\Lambda}=0.7$ are also adopted.
\begin{figure}
\begin{center}
\includegraphics[width=0.63\columnwidth,angle=-90]{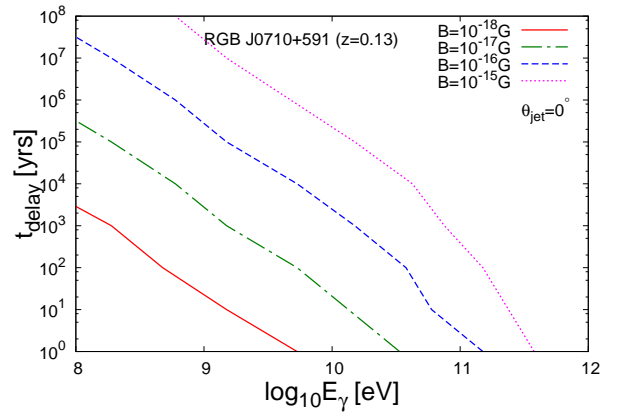}
\caption{The mean time delay incurred by a photon energy flux following its injection at 10$^{14}$~eV from a source at redshift $z=0.13$, for different intervening EGMF, in the range $10^{-18}$-$10^{-15}$~G.}
\label{t_delay}
\end{center}
\end{figure}

In the calculations throughout this work, the EGMF are assumed to be 
describable by patches of coherent (uniform) magnetic field, each patch being $\lambda_B=1$~Mpc in size\footnote{We have verified that the derived limits on EGMF do not depend on $\lambda_B$ for $\lambda_B\ge 1$~Mpc by comparing the results of calculation of the cascade for $\lambda_B=1$~Mpc with those for $\lambda_B=30$~Mpc.}. Different 
magnetic patches have their fields orientations chosen completely independently (randomly) of each other. Such a
description amounts to assuming that the magnetic field power spectrum places a significant proportion of the
magnetic field energy density at the longest length scales. Since, if this is not the case, the magnetic field
deflections will be somewhat weaker, our results obtained in this work under this assumption can be considered 
conservative.

\begin{figure}
\begin{center}
\includegraphics[width=0.6\columnwidth,angle=0]{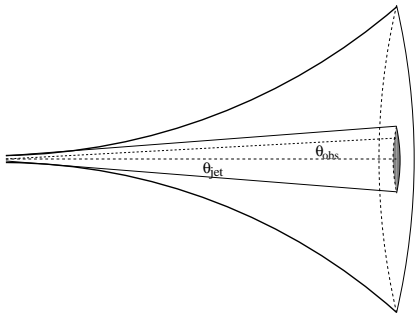}
\caption{A diagram depicting the flaring of the electromagnetic cascade development in the presence of extragalactic magnetic fields. The initial (conical) jet emission whose power feeds the electromagnetic cascade is also shown, with the inner shaded region on the right representing the region filled by the intrinsic cone emission.}
\label{cascade_intrin}
\end{center}
\end{figure}

The presence of a non-negligible magnetic field ($>10^{-20}$~G) introduces
both spatial spreading of the cascade front in angle, away from the initial
beam direction, and the significant growth of the cascade front depth 
(arrival time spread \citep{Plaga:1995}). Spatial and time decomposition of the full cascade spectrum arising at non-zero EGMF is shown
in the top and bottom panels of Fig.~\ref{angle_time_cuts}, respectively.

\begin{figure*}
\includegraphics[width=0.28\linewidth,angle=-90]{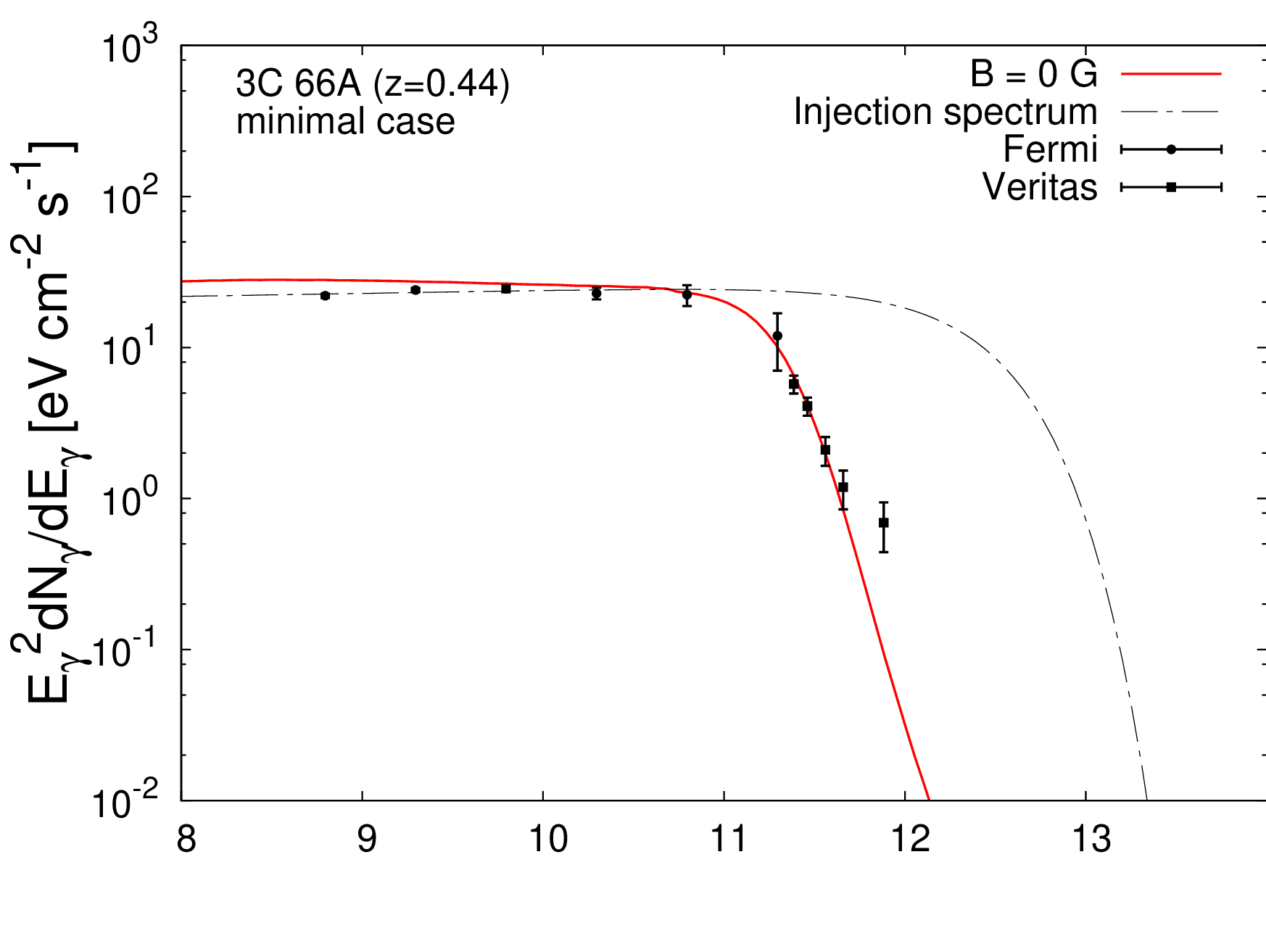}
\includegraphics[width=0.28\linewidth,angle=-90]{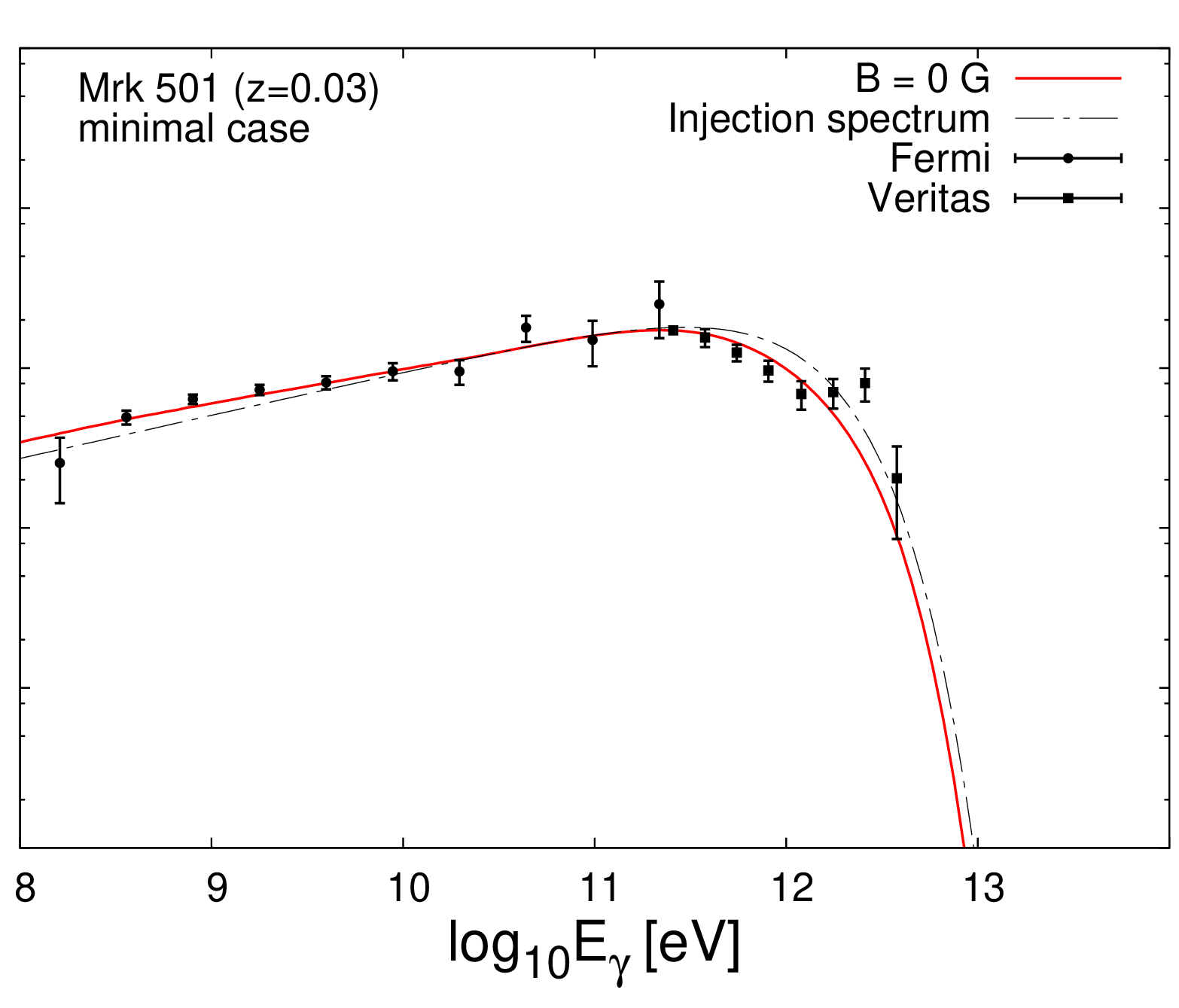}
\includegraphics[width=0.28\linewidth,angle=-90]{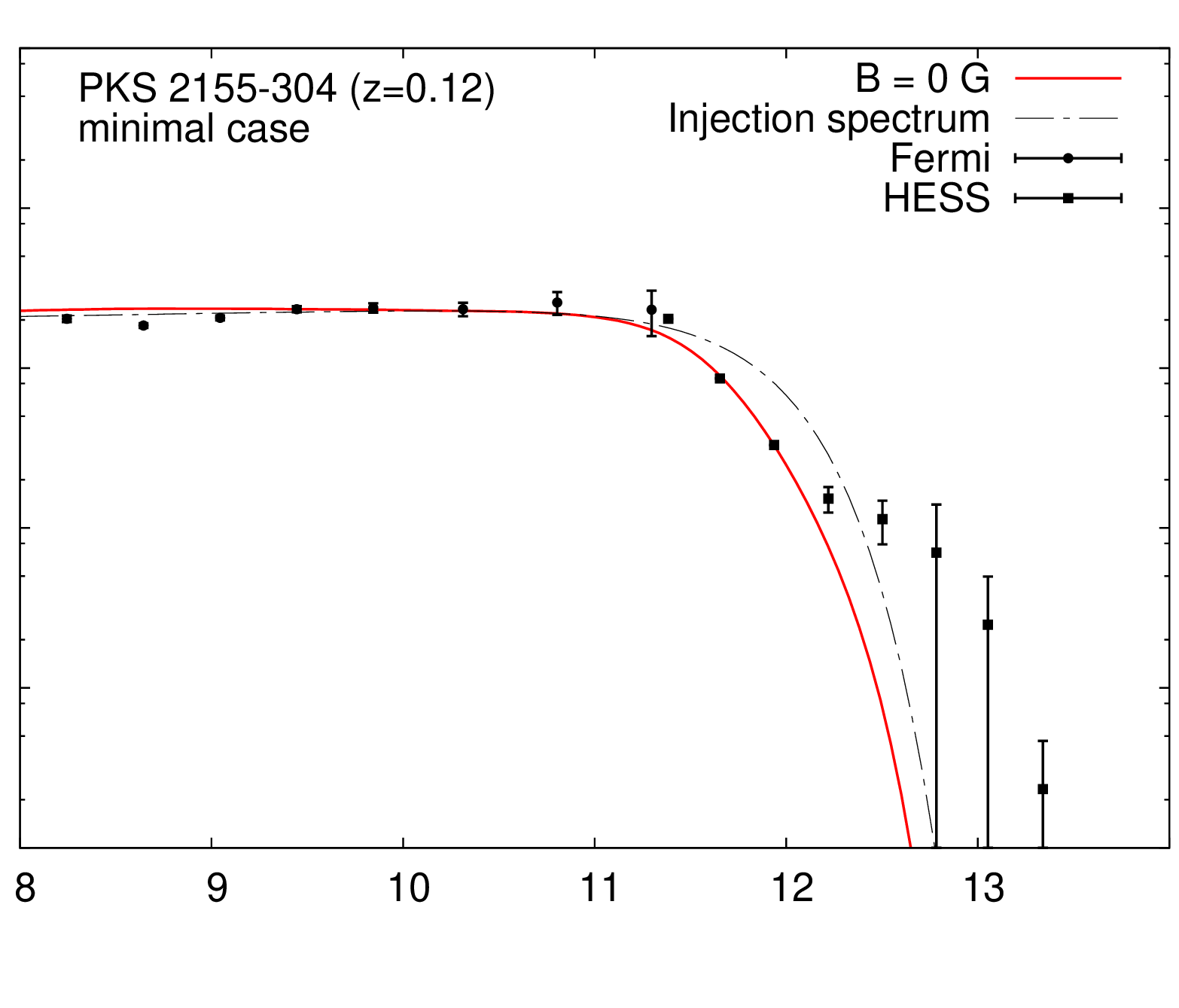}
\caption{The arriving photon energy flux, resulting from
the ``minimal cascade'' model for Mrk 501 (top), PKS 2155-304 (middle) and 3C~66A (bottom). Long-dash-dotted line shows the intrinsic source spectra. Red solid lines show the spectra with account of the cascade component.}
\label{mrk-pks-3C66}
\end{figure*}

For the calculation shown in the top panel of Fig.~\ref{angle_time_cuts}, an observer is placed at an off-axis angle $\theta_{obs}$ with respect to the gamma-ray beam, at a distance $D$ from the source. The observer detects only photons incident on the sphere of the radius $R=D$ at small incidence angles $\theta\le \theta_{\rm PSF}$ with respect to the  normal to the sphere. $\theta_{\rm PSF}$ corresponds to the point spread function of observer's telescope.   
An approximation of $\theta_{\rm PSF}$ for 
the {\it Fermi}/LAT instrument, valid for photon energies in the range 30~MeV to 
300~GeV 
\begin{eqnarray}
\theta_{\rm PSF}\approx 1.7^{\circ} \left(E_{\gamma,{\rm GeV}}\right)^{-0.74}\left[1+\left(\frac{E_{\gamma,{\rm GeV}}}{15}\right)^{2}\right]^{0.37},
\end{eqnarray}
where $E_{\gamma, {\rm GeV}}$ is the photon energy in GeV, is used for the purposes of our calculations\footnote{See http://www-glast.slac.stanford.edu/software/IS/\\ glast\_lat\_performance.htm}. 
Bin widths of $0.1^{\circ}$ (i.e. photons are collected from a ring $\theta_{obs}\pm 0.1^\circ$) are used for the observer position.
In the bottom panel of Fig.~\ref{angle_time_cuts}, only photons arriving 
with a particular delay time, relative to that of the propagation time of the direct gamma-ray photons 
from the source, are shown. The width of the time bins is half a decade around the reference $t_{\rm delay}$ value.
\begin{figure*}
\includegraphics[width=0.28\linewidth,angle=-90]{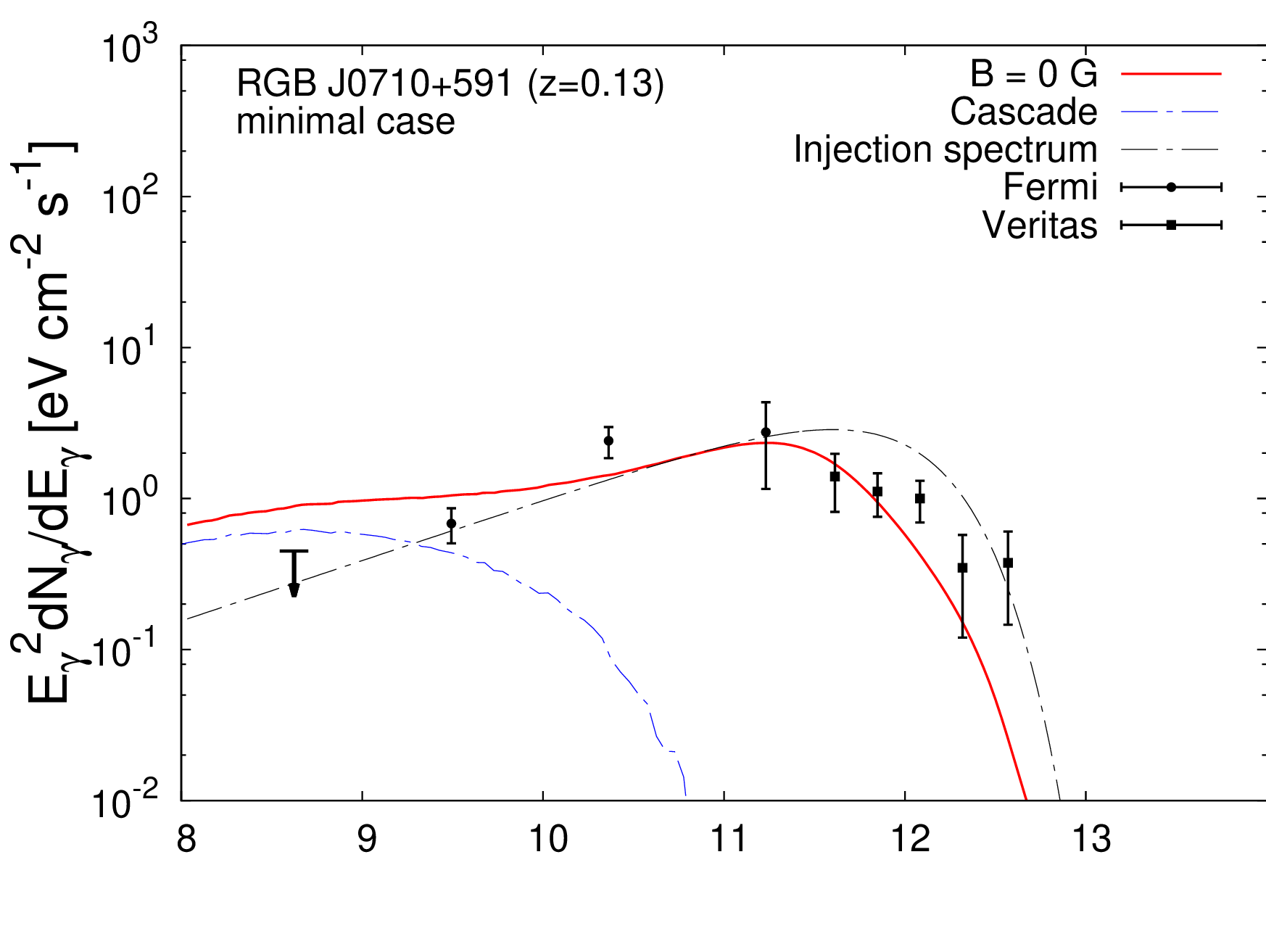}
\includegraphics[width=0.28\linewidth,angle=-90]{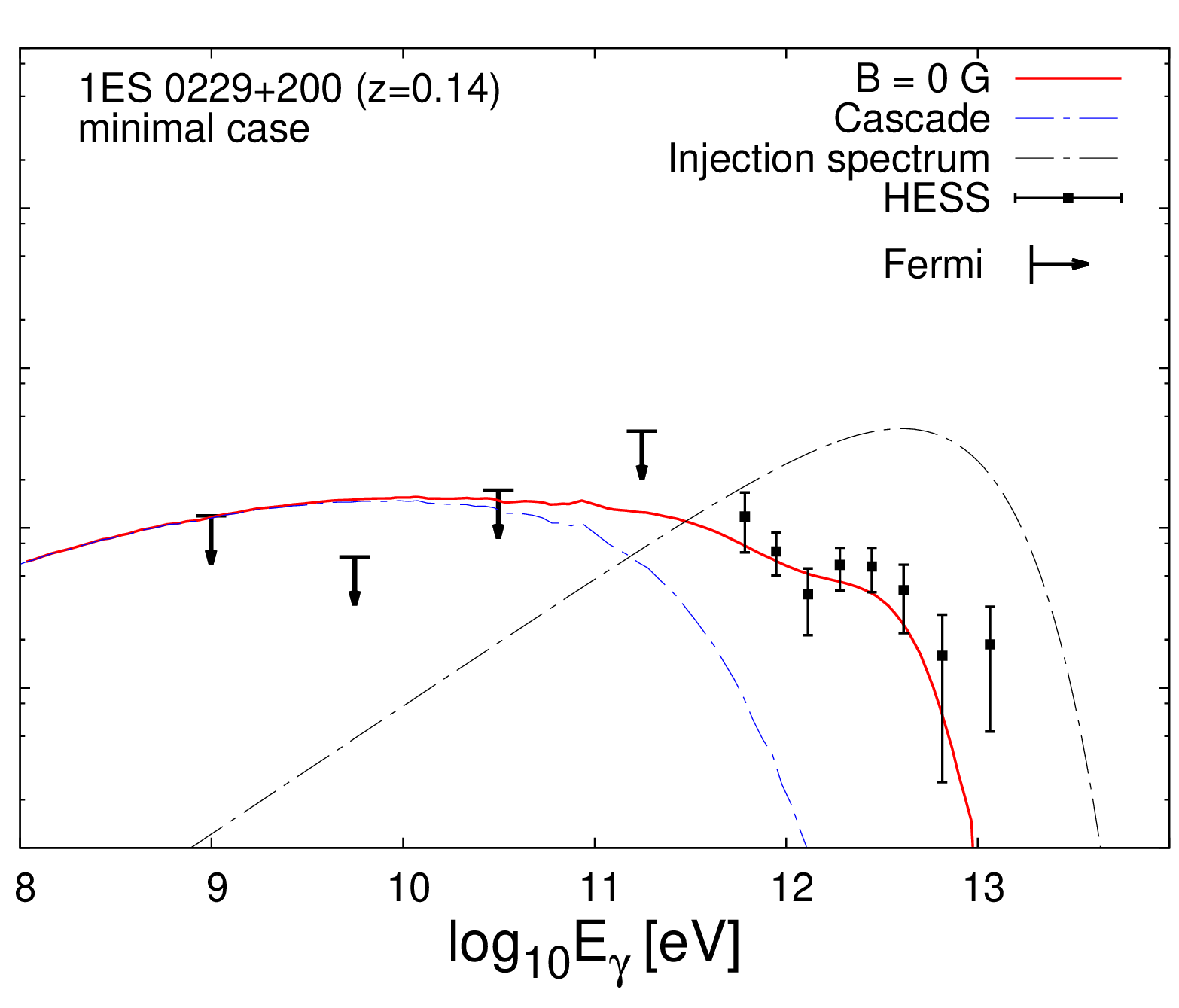}
\includegraphics[width=0.28\linewidth,angle=-90]{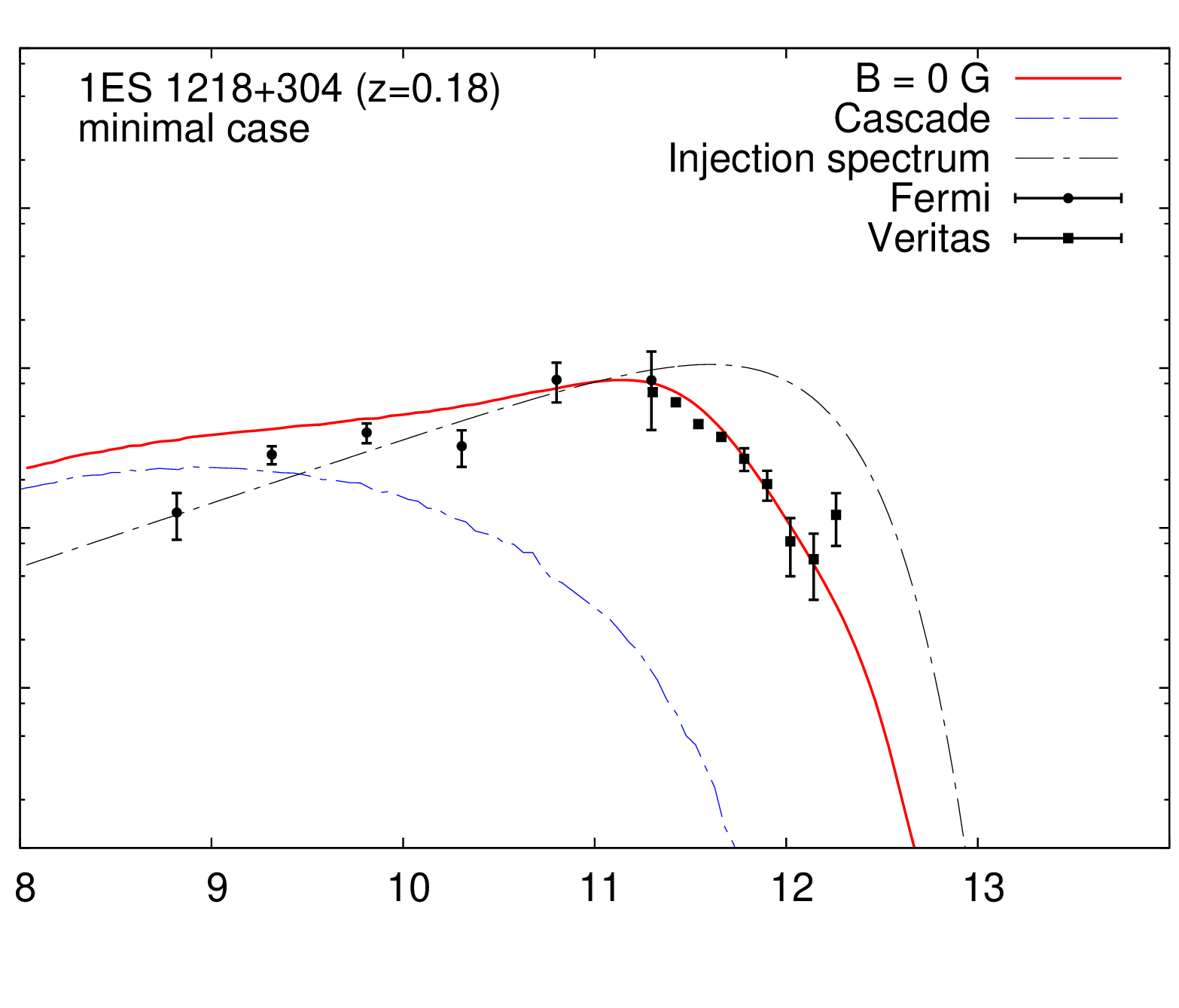}
\includegraphics[width=0.28\linewidth,angle=-90]{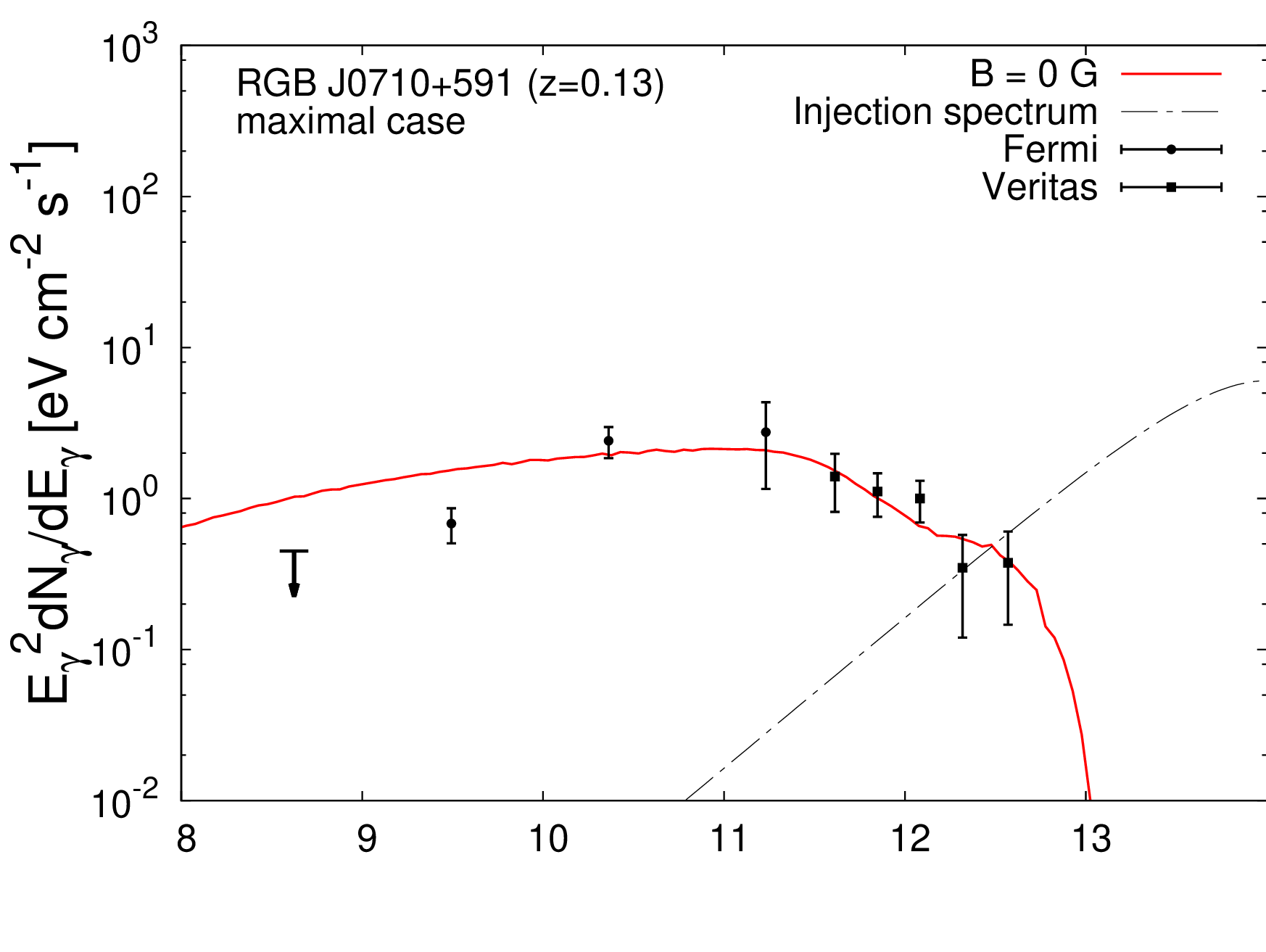}
\includegraphics[width=0.28\linewidth,angle=-90]{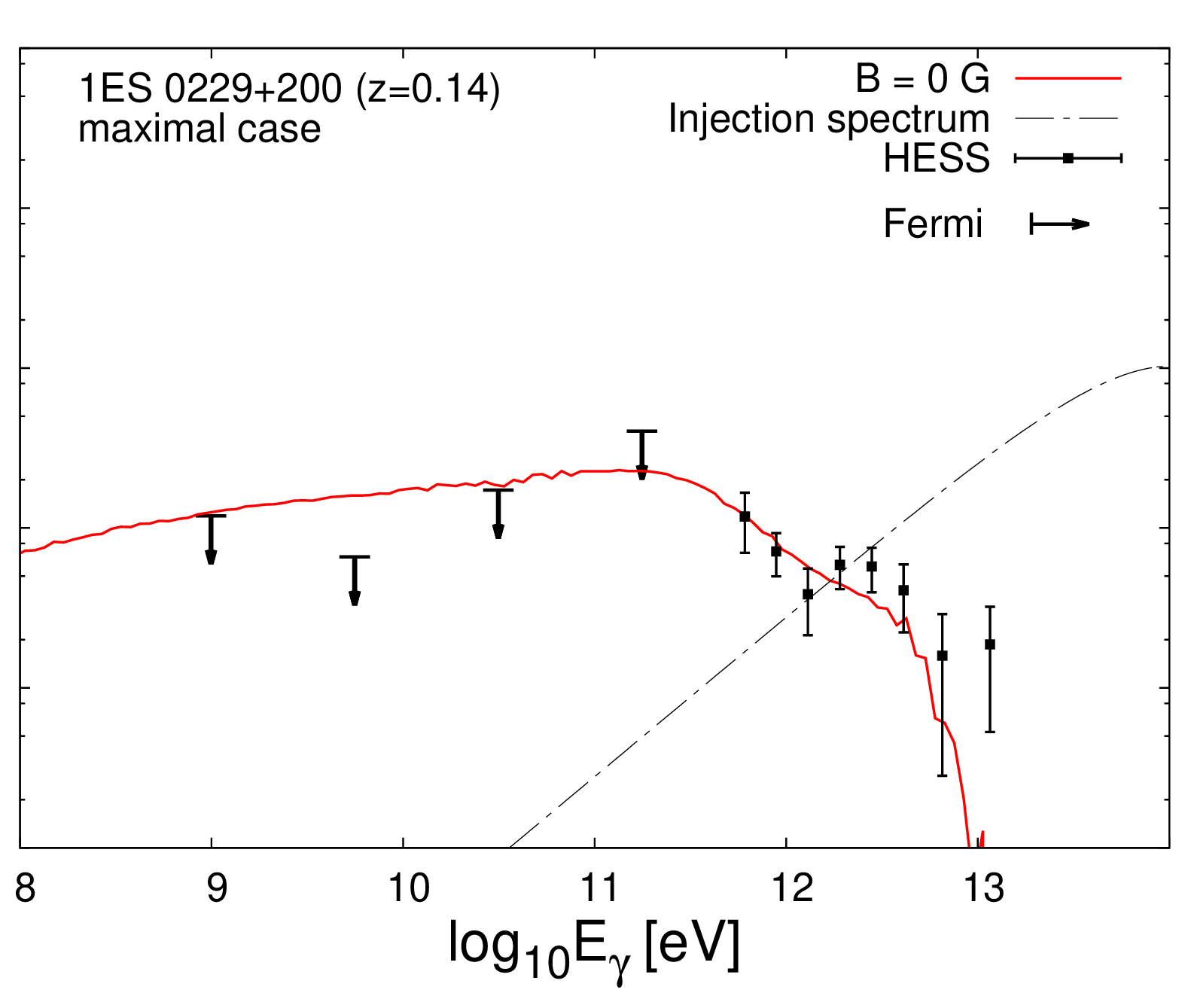}
\includegraphics[width=0.28\linewidth,angle=-90]{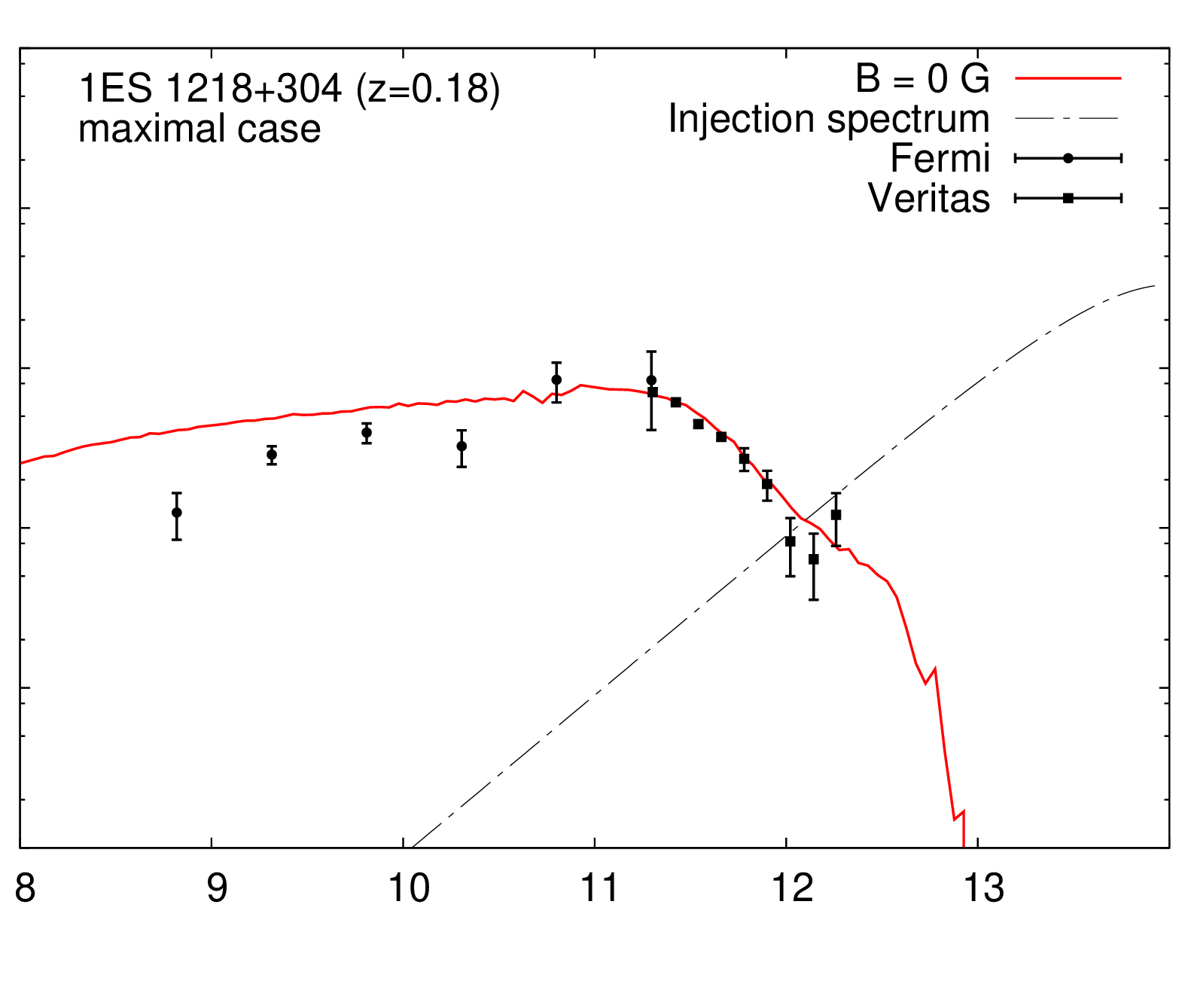}
\caption{Top: The arriving photon energy flux, resulting from
the ``minimal cascade'' model for RGB~J0710+591 (left), 1ES~0229+200 (center) and 1ES~1218+304 (right). Fermi upper bounds are at the 99\% confidence level. Notations are the same as in Fig. \ref{mrk-pks-3C66}. Blue dash-dotted line shows the cascade component of the spectrum. Bottom: the results for the ``maximal cascade'' model.}
\label{minimal-maximal}
\end{figure*}

Fig. \ref{t_delay} shows the time delay of the cascade photons as a function of the photon 
energy for different magnetic field strengths. Coarse bin widths of 1 per logarithmic decade in time are 
used for time delay binning. One can see that the time delay decreases proportionally $\sim E_{\gamma}^{-2.5}$ or  $\sim E_{\gamma}^{-2}$,  with the increasing photon energy $E_{\gamma}$, in agreement with the analytical calculations \citep{neronov09}.  Change of the the slope of the energy dependence of the time delay at high energies is related to the energy dependence of the mean free path of the primary $\gamma$-rays.

\subsection{Jet Opening Angle}

A realistic description of jet cascade effects in an EGMF should take
into account the intrinsic spread in angles of the injected gamma-rays, within the opening angle of the blazar jet, $\theta_{\rm jet}$ as well as the spread in initial energies of gamma-rays, $E_{0}$.

To take into account the finite opening angle of the blazar jet we use the same procedure as described in \cite{elyiv09},  utilizing zero jet width results. Using a random remapping 
of the arriving particles' position and velocity vectors on the arrival surface, 
of the zero jet width results, the effect of the jet width on the cascade results 
is obtained. Fig.~\ref{cascade_intrin} illustrates the geometrical set up of our Monte Carlo calculations. In what follows we fix the jet width to $\theta_{\rm jet}=5^\circ$, which is typical opening angle of the jets observed in radio galaxies and is the angle $\theta_{\rm jet}\simeq \Gamma_{\rm jet}^{-1}$ inferred from the typical bulk Lorentz factors of the blazar jets $\Gamma_{\rm jet}\sim 10$ derived from the gamma-ray observations. We (arbitrarily) place the observer  at $\theta_{\rm obs}=2^{\circ}$.

Cascade spectra for the case of a non-zero jet opening angle can be qualitatively understood as the weighted sums of different components of the decomposition of the full cascade spectrum shown in Fig. \ref{angle_time_cuts} (i.e. adding cascade emission produced by primary photons emitted in different directions within $\theta_{\rm jet}$ is equivalent to displacing the observer to a different $\theta_{\rm obs}$ with respect to the primary beam direction). The effect of introducing 
an intrinsic jet opening angle is, therefore, to wash away the small
angle/small time delay information present in the high energy emission.

\subsection{Intrinsic spectra of the blazars}

It is conventionally assumed that the spectra of the blazars measured in the {\it Fermi} energy band are ``intrinsic'' spectra, produced at the source and that only the spectrum above $E_{\gamma}\gtrsim 100$~GeV is strongly influenced by the effects of pair production on the EBL. This assumption does not work, a priori, if the uncertainty of the strength of EGMF is taken into account. In particular, if the EGMF is close to zero, cascade emission can give a significant contribution to the observed source spectrum in the GeV energy band, if the intrinsic spectrum of the source is hard, so that most of the power is initially emitted at energies $E_{\gamma}\gtrsim 100$~GeV. In particular, it is even possible that the cascade component gives a dominant contribution in the source flux in the GeV-TeV band, as was demonstrated for the particular case of the blazar Mrk 501 by \citet{Aharonian:2001cp}. More generally, the observed spectrum in the 0.1~GeV -- 10~TeV energy range, simultaneously measured by {\it Fermi} and by the ground-based Cherenkov telescopes is a superposition of the direct source flux (with an a-priori unknown spectrum) and a cascade flux. The spectrum of the cascade flux can be unambiguously derived from the intrinsic source spectrum, if a particular configuration of EGMF is assumed. Thus, proper modeling of the broad band $\gamma$-ray data of blazars in the 0.1~GeV -- 10~TeV energy range has to be used to derive both the intrinsic source spectra and the properties of EGMF from the model fits to the observed spectra.  

We limit the choice of models for the intrinsic source spectra with the cut-off powerlaw type models
\begin{eqnarray}
\frac{dN_{\gamma}}{dE_{\gamma}}\propto E_{\gamma}^{-\Gamma}\exp\left(-\frac{E_{\gamma}}{E_{\rm cut}}\right),
\label{minimal_model}
\end{eqnarray}
described by two parameters, the photon index $\Gamma$ and the cut-off energy $E_{\rm cut}$. Following \citet{Neronov:2009}, the values of $\Gamma$ and $E_{\rm cut}$ are chosen such that the 
absorbed spectrum, along with the subsequent cascade contribution, give both
good fits to the complete multi-wavelength (GeV and TeV) data set and minimize 
the cascade contribution. Technically, if the cut-off energy $E_{\rm cut}$ is not constrained by the data, the lower bound on $E_{\rm cut}$ is derived from the TeV data and this lower bound is assumed for the derivation of the cascade component of the spectrum. Since this procedure minimizes the fraction of TeV flux absorbed during 
propagation, the fit obtained is labelled as our ``minimal case'' model for 
the cascade contribution.

In addition to the powerlaw type spectra we also consider a qualitatively different model for the intrinsic source spectra,  which results in the dominant cascade contribution in the GeV band. We call this case the ``maximal'' cascade model. As is mentioned above, as soon as the bulk of the initial source power is injected at energies far above the threshold for pair production on the EBL, the resulting shape of the cascade component of the spectrum almost does not depend on the details of the intrinsic source spectrum. Taking this into account, we arbitrarily fix the intrinsic photon index to  $\Gamma=1$ and the cut-off energy to $E_{\rm cut}=100$~TeV. The results for the observed GeV-TeV band spectrum in the maximal cascade model are similar to those derived by \citet{Aharonian:2001cp} for the case of negligible EGMF.

\section{Results}
\label{Results}

\subsection{Minimal and maximal cascade model fits to the spectra}
 
From the combined broad 0.1~GeV -- 10~TeV band gamma-ray spectra of the sources listed in Table \ref{sources_list},  we find that in general, the high energy end of the {\it Fermi}/LAT spectra at $E_{\gamma}\sim 100$~GeV perfectly matches the low energy end of the spectra obtained with the ground based gamma-ray telescopes. This indicates that no additional ``intercalibration factors'' between different instruments are needed in the modeling of the source spectra. 

The observed 0.1-100~GeV band spectra of Mrk~501, PKS~2155-304 and 3C~66A have photon indices close to $\Gamma=2$ with almost no curvature of the spectrum over the entire energy range. Such a ``flat'' spectral energy distribution is difficult to obtain in the ``maximal cascade'' model. Although the ``universal'' cascade spectrum 
has average photon index close to $\Gamma=2$ in the 0.1-100~GeV band, its deviation from the pure powerlaw behavior is significant, so that this model is in contradiction with the observed spectrum for these blazars.
This suggests that the dominant contribution to the observed source spectrum in the cases of Mrk~501, PKS~2155-304 and 3C~66A comes from direct gamma-rays from the source as shown in Fig. \ref{mrk-pks-3C66}.

\begin{figure}
\begin{center}
\includegraphics[height=0.8\columnwidth,angle=-90]{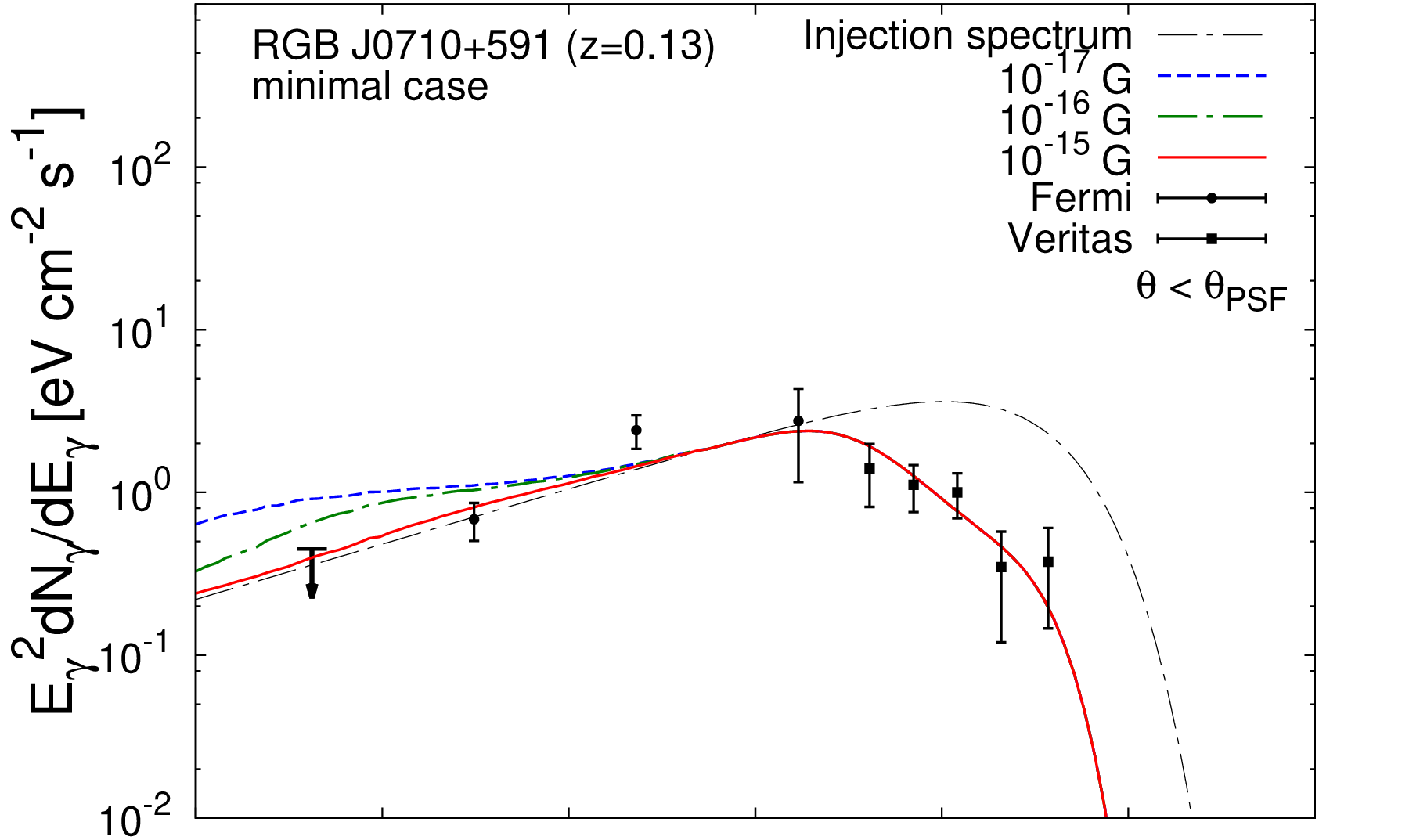}
\includegraphics[height=0.8\columnwidth,angle=-90]{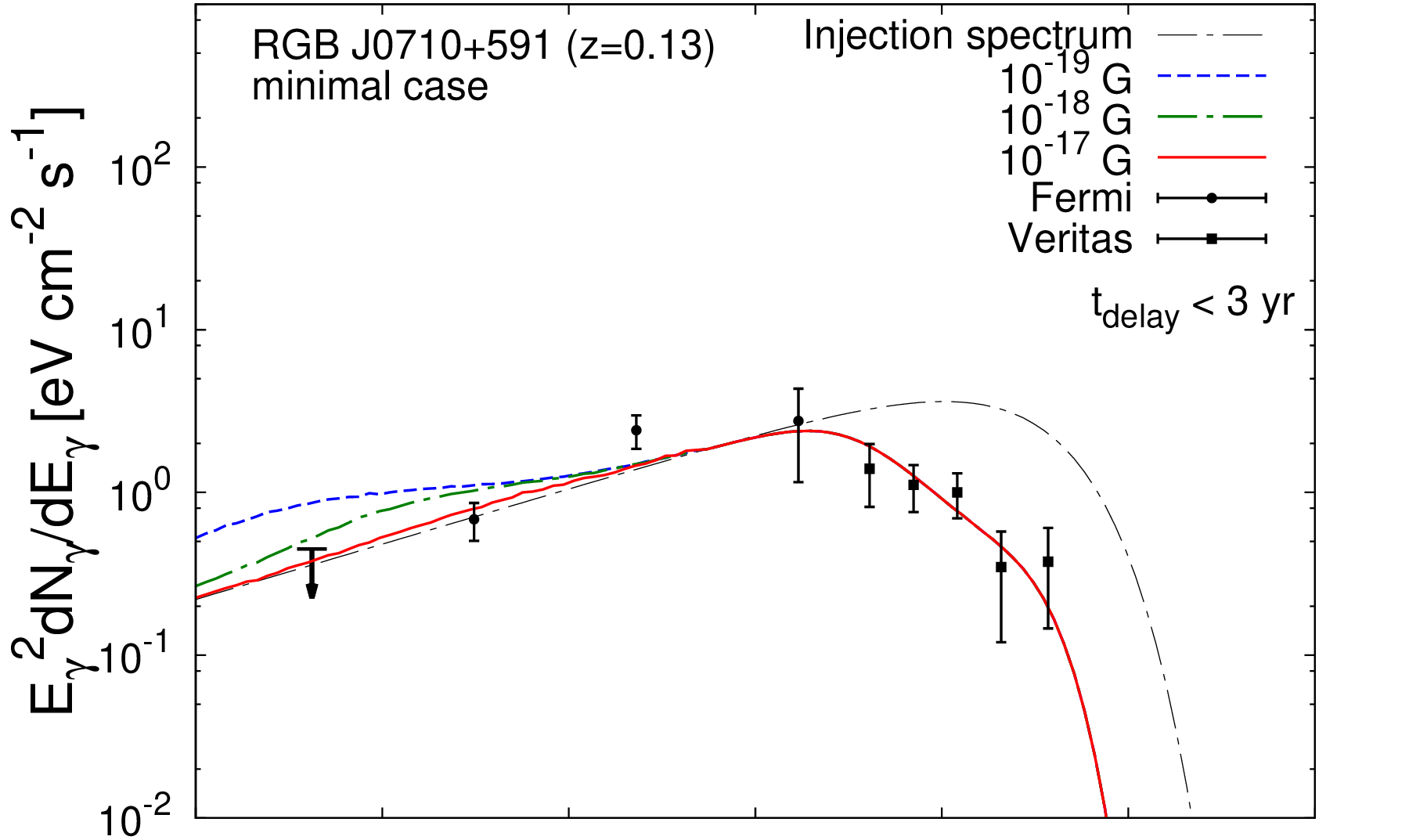}
\includegraphics[height=0.8\columnwidth,angle=-90]{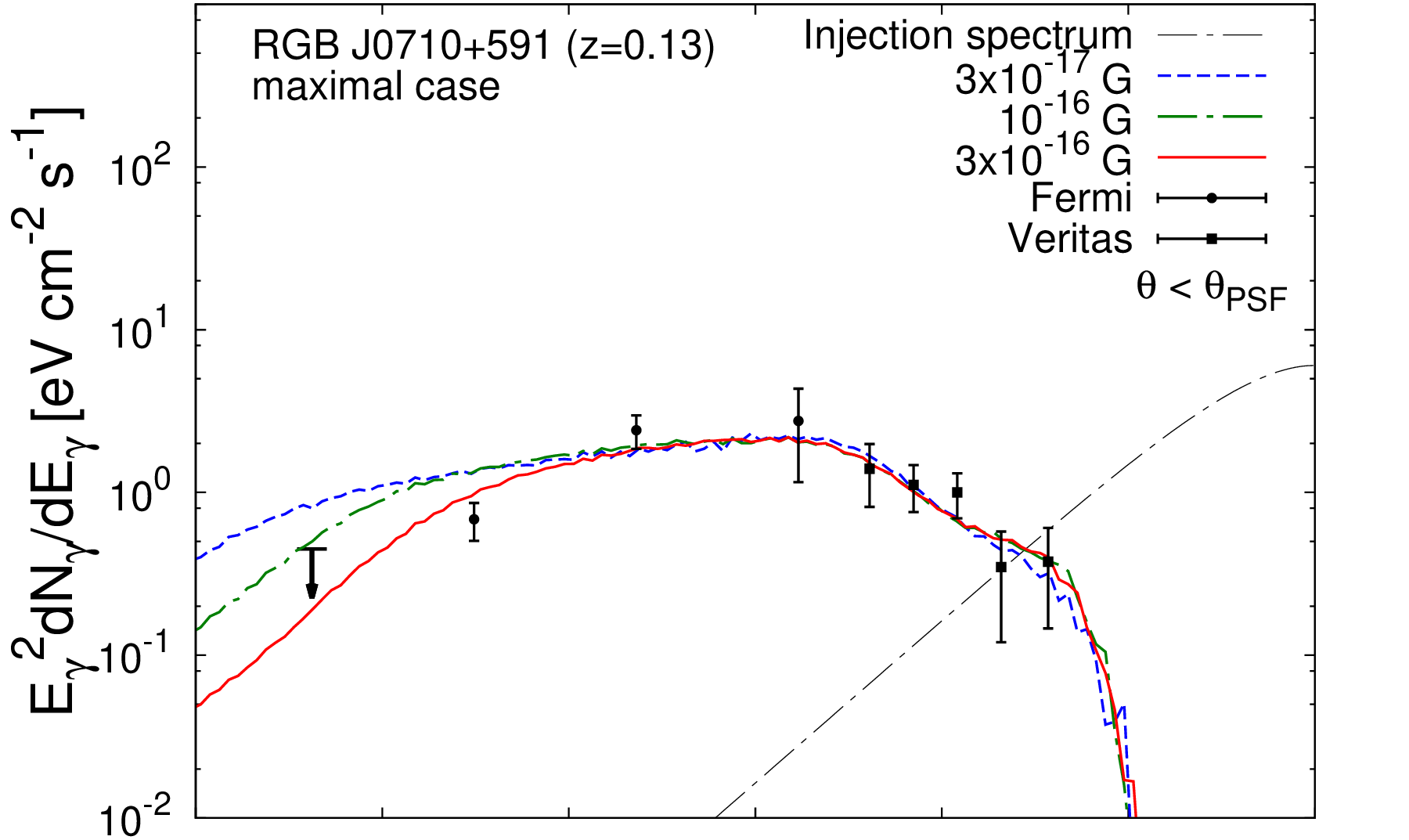}
\includegraphics[height=0.8\columnwidth,angle=-90]{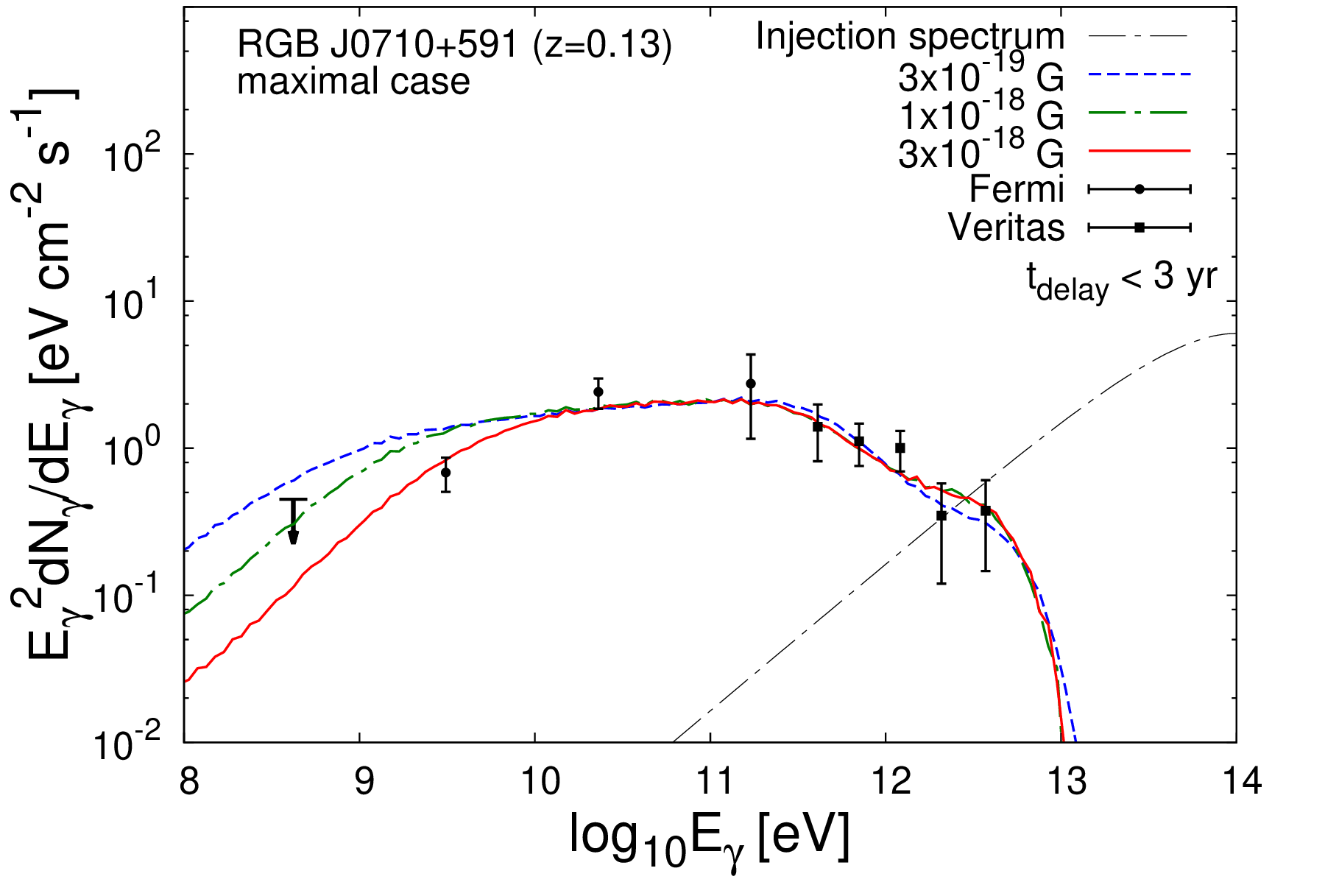}
\caption{The effects of the presence of an EGMF on the arriving cascade from the blazar RGB~J0710+591. Two top panels show the results for the case of the ``minimal cascade'' model of the $\gamma$-ray spectrum, for the two possibilities of suppression of the cascade flux via extension of the cascade source and time delay of the cascade emission. Two bottom panles show the results for the ``maximal cascade'' model.}
\label{suppression_RGB}
\end{center}
\end{figure}

Addition of the cascade contribution to the intrinsic emission component in Mrk~501, PKS~2155-304 and 3C~66A leads only to a moderate modification of the source spectrum in the GeV energy range. Slight adjustment of the intrinsic spectrum photon index is sufficient to make the observed spectrum consistent with the model even for the case of zero EGMF strength. The reason for the sub-dominance of the cascade contribution in these sources is the softness of the intrinsic source spectrum ($\Gamma\simeq 2$). The fraction of the source power absorbed on the way from the source to Earth is at most comparable to the primary source power at energies $E_{\gamma}<100$~GeV, so that cascade emission can not dominate over the intrinsic source power.

\begin{figure}
\begin{center}
\includegraphics[height=0.8\columnwidth,angle=-90]{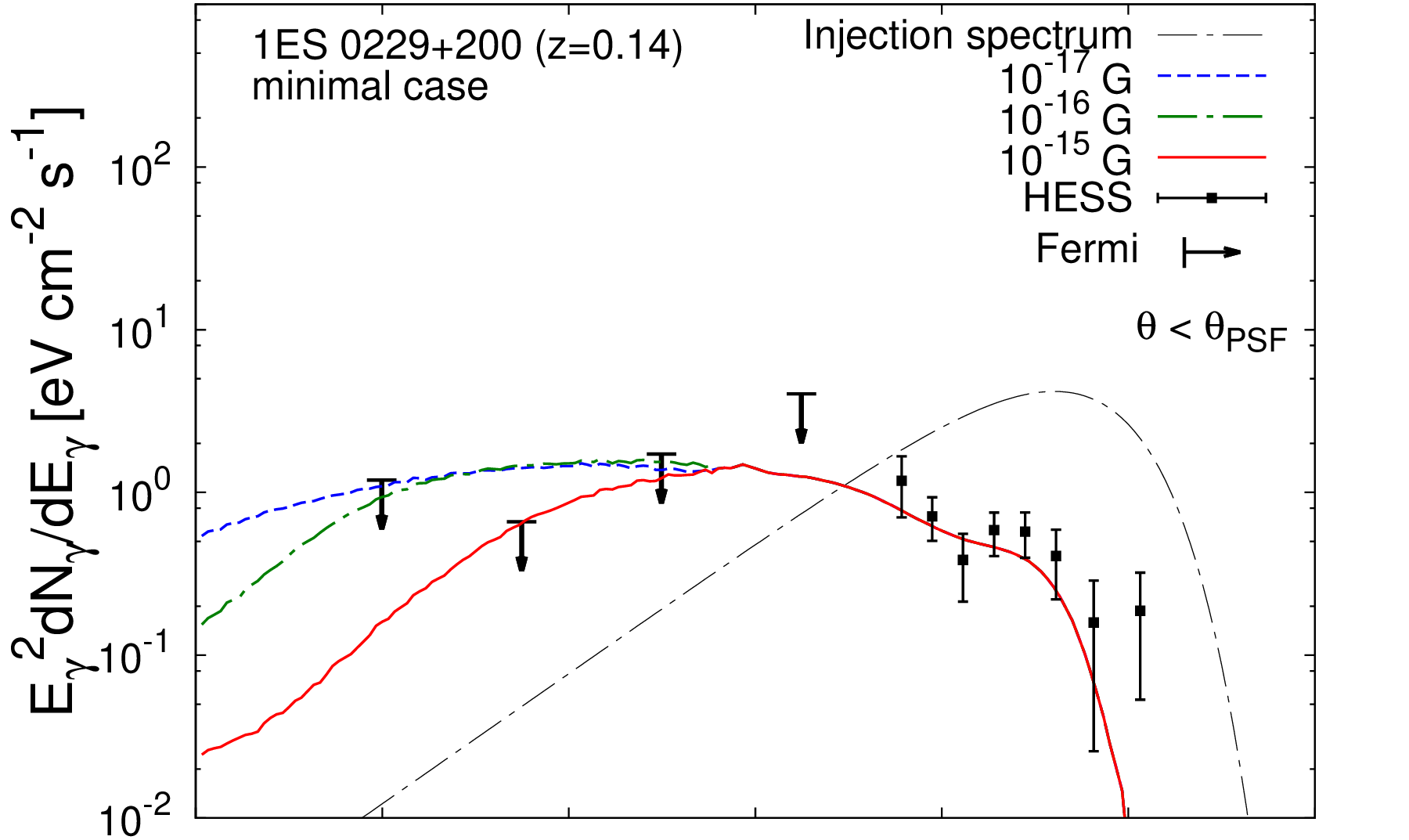}
\includegraphics[height=0.8\columnwidth,angle=-90]{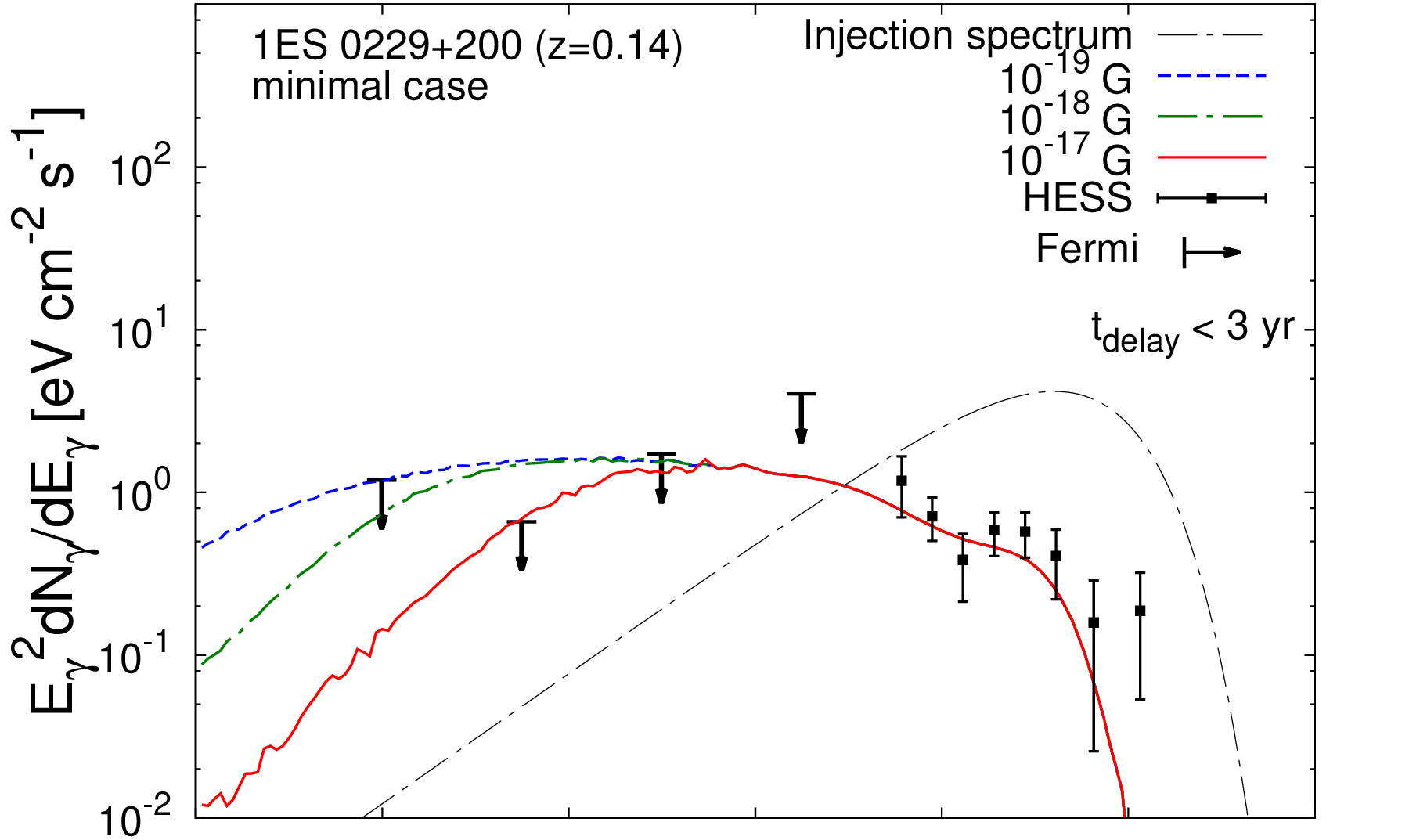}\\
\includegraphics[height=0.8\columnwidth,angle=-90]{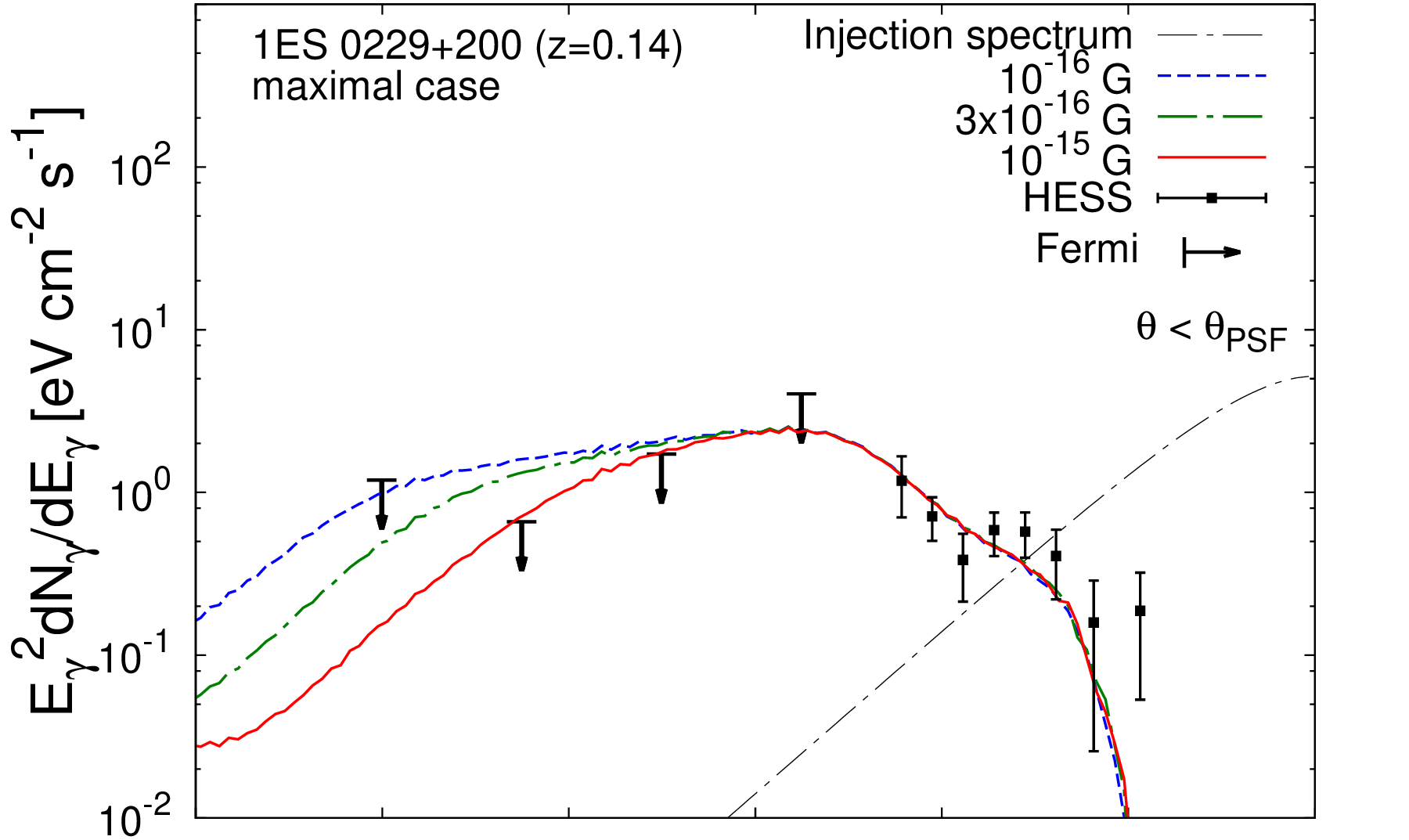}
\includegraphics[height=0.8\columnwidth,angle=-90]{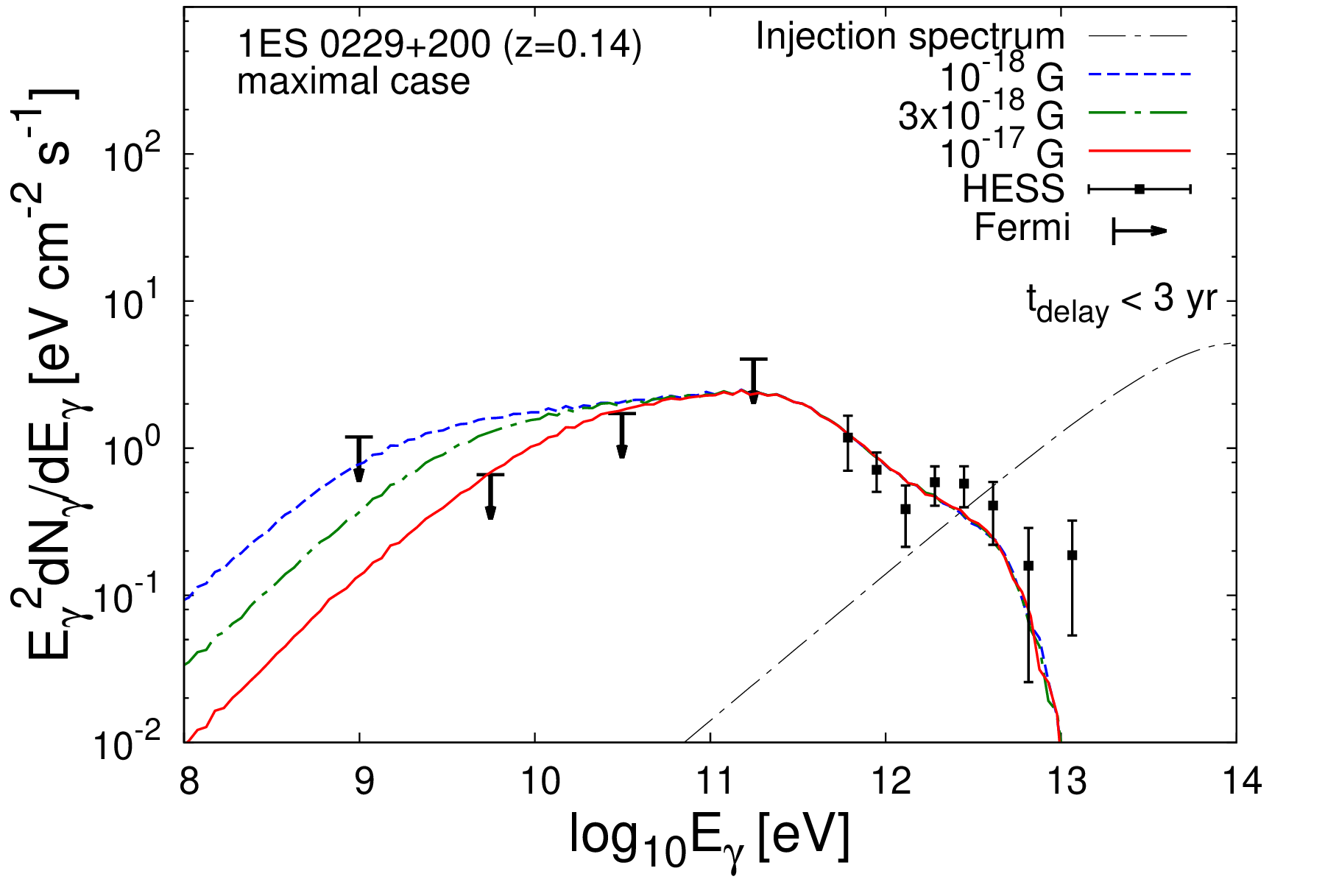}
\caption{Same as in Fig. \ref{suppression_RGB} but for 1ES~0229-3+200. }
\label{suppression_1ES_0229}
\end{center}
\end{figure}

The situation is different, however, for the case of RGB~J0710+591, 1ES~0229+200, and 1ES~1218+304. The intrinsic spectra  of these sources are harder than $\Gamma= 2$. The hard intrinsic spectrum of the source leads to a large power output at energies above 100~GeV, even in the case when a possible high energy cut-off in the source spectrum is taken into account. In the ``minimal'' model shown in Fig. \ref{minimal-maximal}, the intrinsic source power output in the TeV energy band is an order of magnitude higher than that in the GeV band. A significant fraction of the of the intrinsic source power in the TeV band is absorbed through pair production on the EBL. This absorbed power is re-emitted in the GeV band, so that the cascade emission in the 0.1-10~GeV range dominates over the intrinsic source emission (solid red curve in Fig.\ref{minimal-maximal}).

The fit of the model (\ref{minimal_model}) alone to the observed spectra result in reduced $\chi^2$ close to 1 both for RGB~J0710+591 and 1ES~1218+304. For 1ES~0229+200 only an upper bound on the source flux in the GeV band could be derived from {\it Fermi} data. Account  of the cascade contribution to the source spectra, which has to be present for the case of zero EGMF, violates the upper bound on the GeV band source flux in the case of 1ES~0229+200. In the case of RGB~J0710+591 and 1ES~1218+304 account of the cascade contribution leads to the worsening of the fit to the spectrum from  $\chi^2=8$ (7 d.o.f.) to $\chi^2=18$ (RGB~J0710+591)  and from $\chi^2=17$ (13 d.o.f.) to $\chi^2=57$ (1ES~1218+304). This implies that the model with cascade component calculated under the assumption of zero EGMF is ruled out at $98.8\%$ and $>99.99\%$ confidence levels for RGB~J0710+591 and 1ES~1218+304,  respectively.

In the case of 1ES~0229+200, the minimal model parameters suffer from a large uncertainty: both the cut-off energy $E_{\rm cut}$ and the spectral slope $\Gamma$ could not be derived from the data. Following \citet{Neronov:2009} we fix the model parameters for this source in such a way that the total flux of the cascade component of the source spectrum is minimized. This is achieved with a very hard value of $\Gamma\sim 1.2$ and relatively high cut-off energy $E_{\rm cut}\sim 5$~TeV, so that the ``minimal'' and ``maximal'' models for 1ES~0229+200 are not very different.

The problem of inconsistency of the predicted cascade flux with {\it Fermi} measurements in the GeV band  is encountered in the ``maximal'' model shown in Fig.~\ref{minimal-maximal}. Assuming that the cascade emission dominates over the direct source emission in the TeV band, one finds that the expected cascade flux level in the 0.1-10~GeV band is higher than the observed source flux, if EGMF strength is $B=0$ (solid red curves in Fig.~\ref{minimal-maximal}).

\subsection{Implications for EGMF}

\begin{figure}
\begin{center}
\includegraphics[height=0.8\columnwidth,angle=-90]{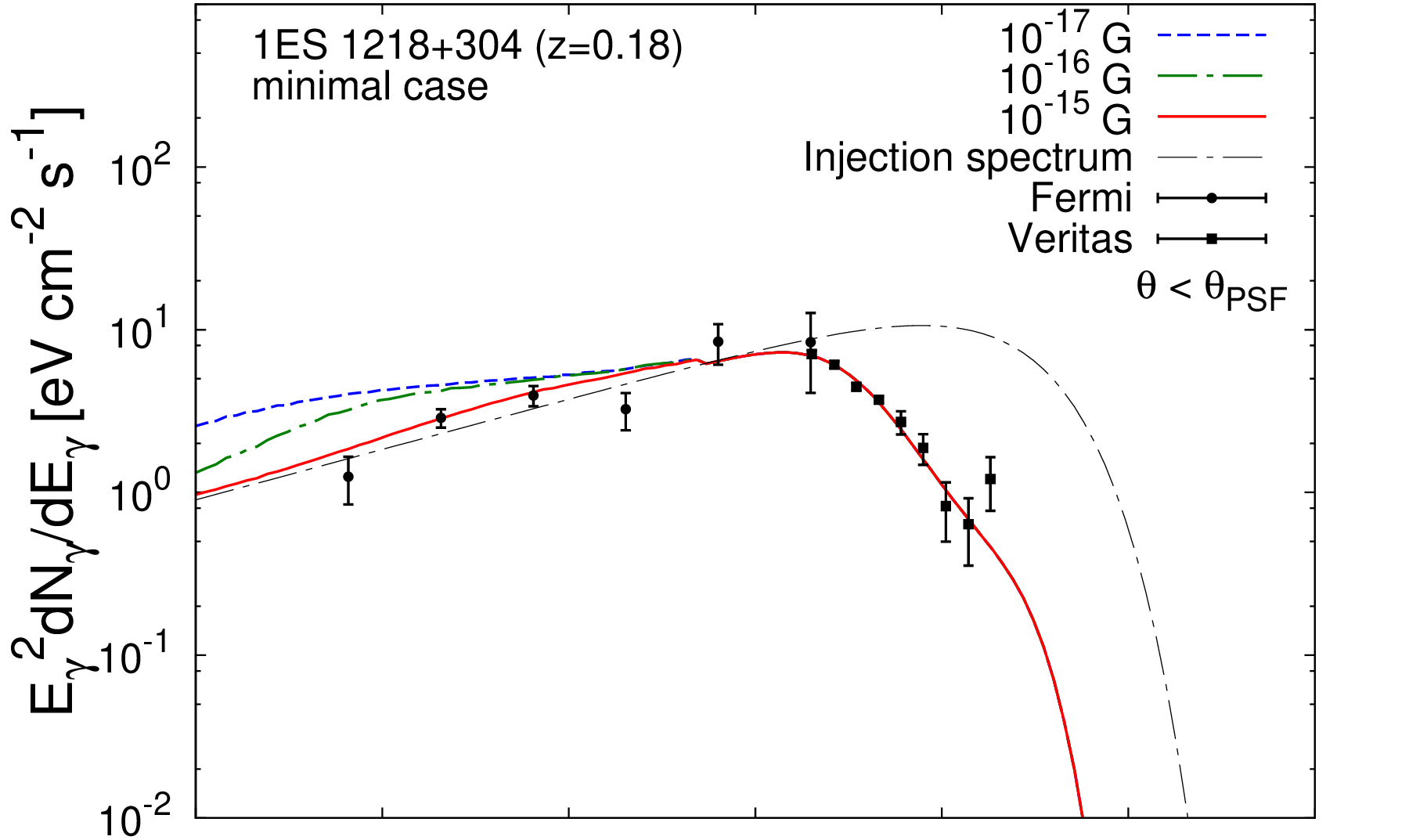}
\includegraphics[height=0.8\columnwidth,angle=-90]{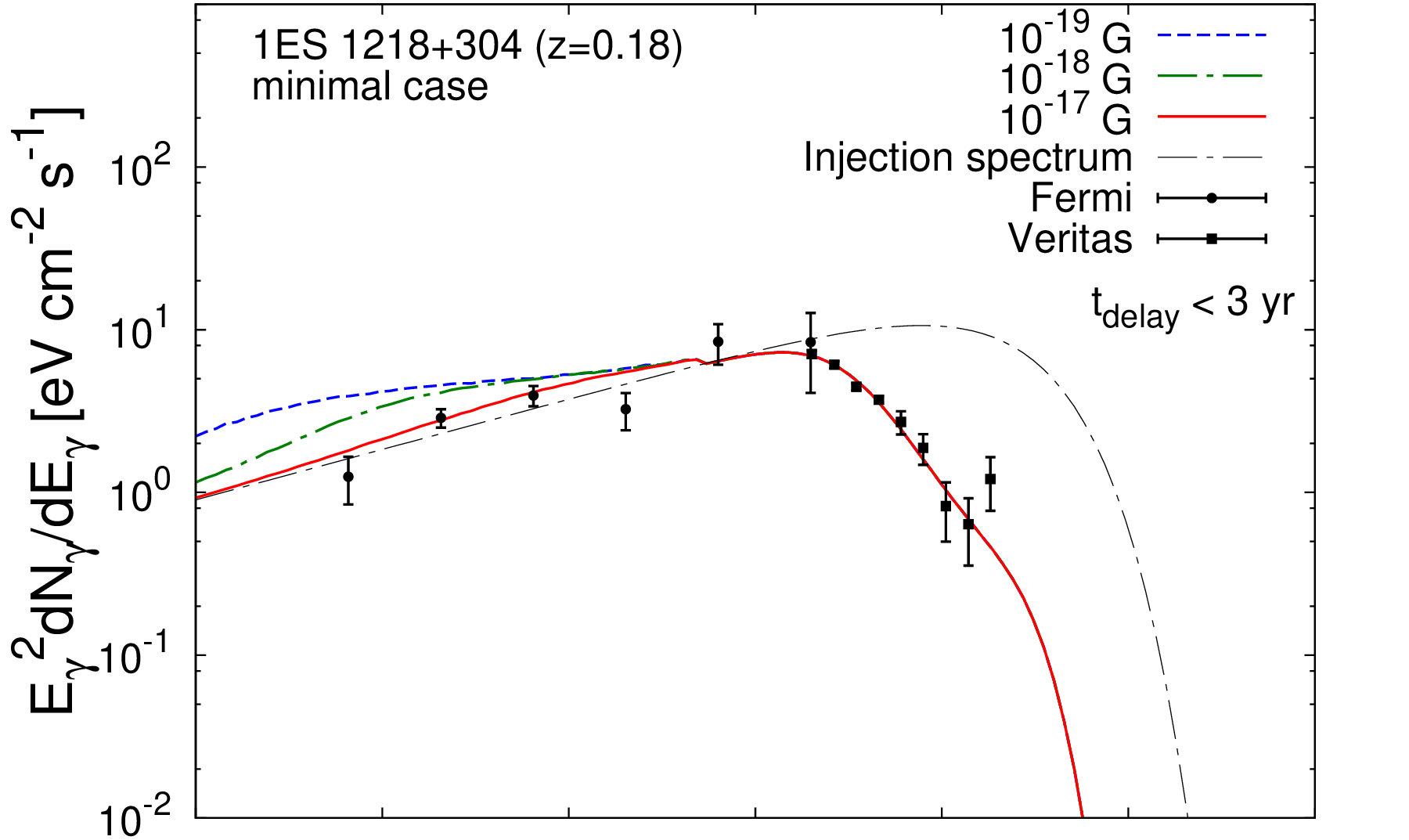}
\includegraphics[height=0.8\columnwidth,angle=-90]{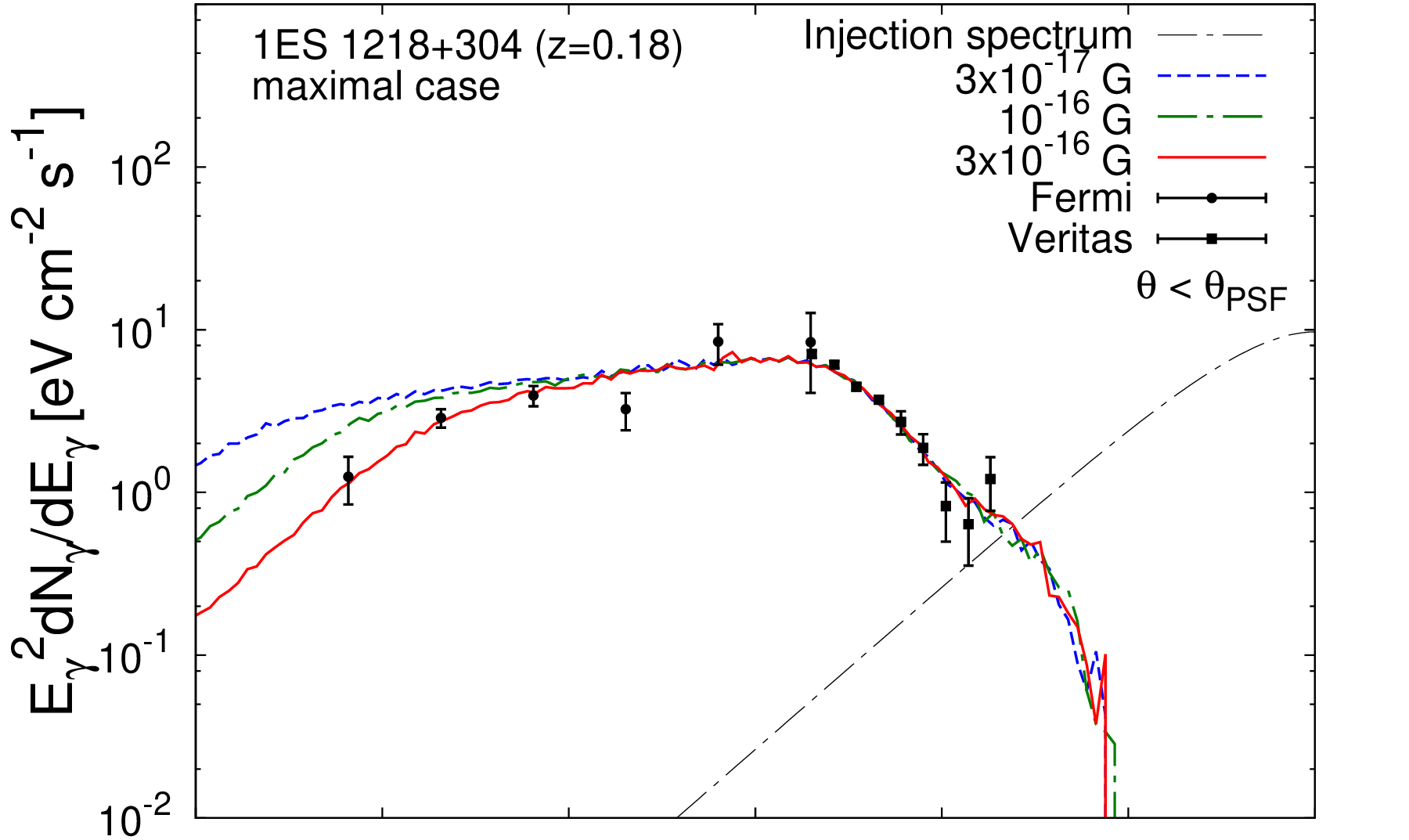}
\includegraphics[height=0.8\columnwidth,angle=-90]{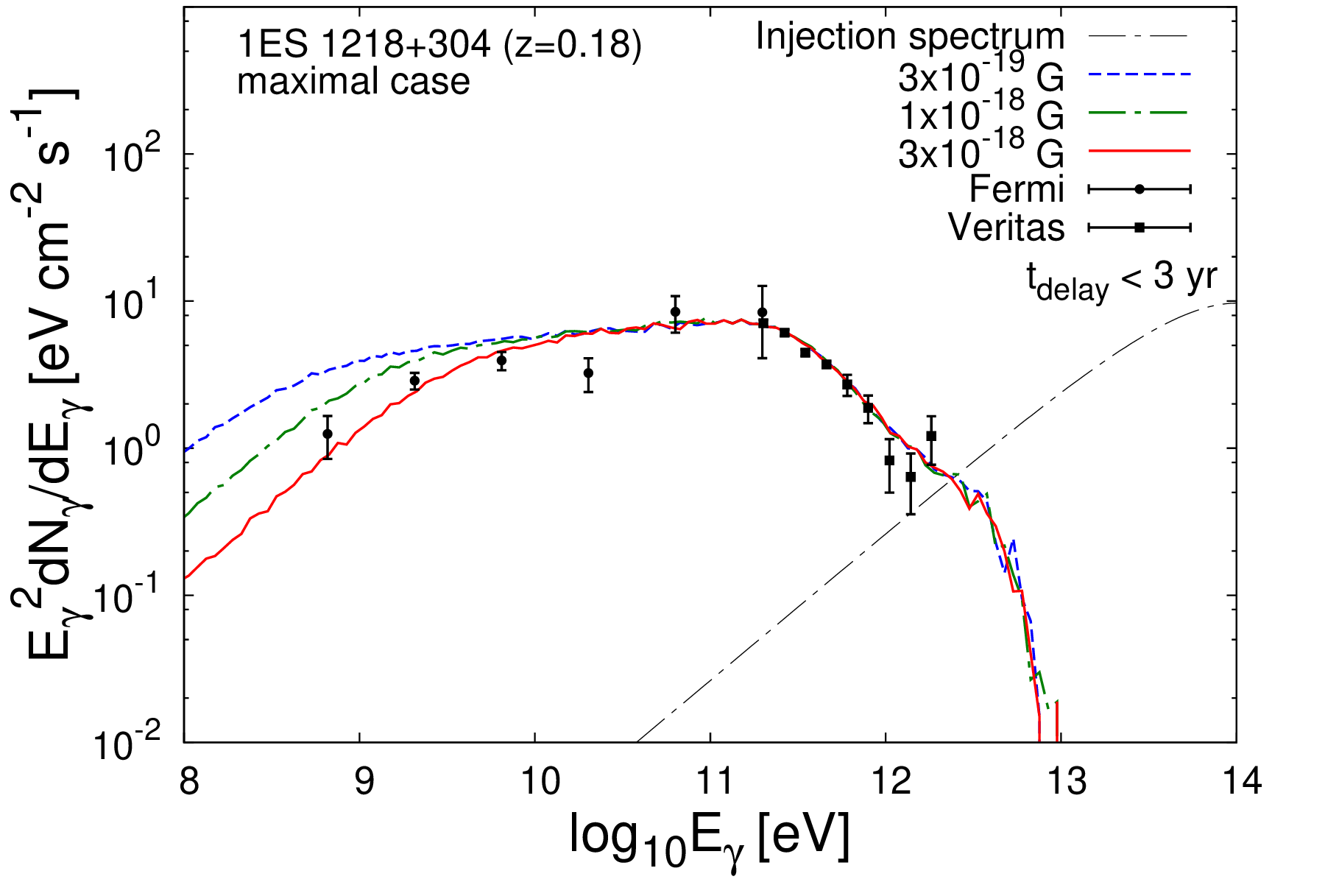}
\caption{Same as in Fig. \ref{suppression_RGB} but for 1ES~1218+304.}
\label{suppression_1ES_1218}
\end{center}
\end{figure}

The fits to the combined GeV-TeV band spectra found in the minimal and maximal models for the cascade contribution under $B=0$ assumption are shown in Fig.~\ref{minimal-maximal} 
for RGB~J0710+591, 1ES~0229+200, and 1ES~1218+304. For these sources, both minimal and maximal models, calculated under the assumption of zero EGMF, are not acceptable. The main source of discrepancy between the model predictions and the data is an over prediction of low energy flux, $E_{\gamma}\lesssim 1$~GeV, due to the presence of the cascade contribution to the total source flux.

\begin{figure}
\begin{center}
\includegraphics[width=\linewidth]{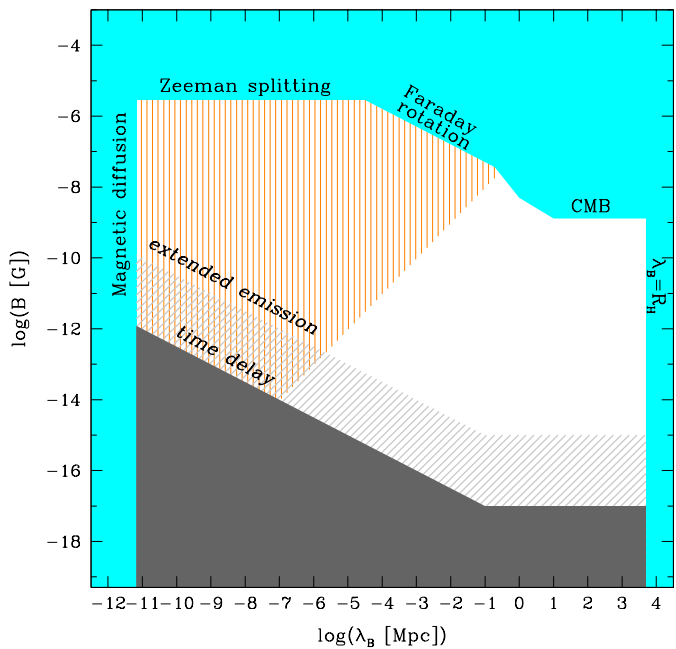}
\caption{Bounds on magnetic field derived from the simultaneous GeV-TeV data. Blue shaded regions show the previously known bounds on $B$ and $\lambda_B$, summarized by \citet{neronov09}. Orange shading shows the upper bound on $B, \lambda_B$  which could be generated before the epoch of recombination, derived by \citet{jedamzik04}.  }
\label{exclusion_time}
\end{center}
\end{figure}

As discussed in section~\ref{Monte_Carlo}, a decrease in the cascade contribution to the arriving flux at low energies can be achieved through the introduction of non-zero magnetic field in the cascade development region.  The effect of a non-zero magnetic field on the cascade component is shown in Figs. \ref{suppression_RGB}, \ref{suppression_1ES_0229}, and \ref{suppression_1ES_1218} for the cases of suppression of the cascade flux due to the large extension of the cascade source and due to the time delay of the cascade signal.

For the case of suppression due to extended nature of the cascade source, the presence of magnetic field modifies the cascade spectrum at GeV energies only if the magnetic field strength is $B\gtrsim 10^{-16}$~G, for the case of RGB~J0710+591, 1ES~0229+200, and 1ES~1218+304. The minimal magnetic field strengths needed to make the model source spectra consistent with the data can vary between $10^{-16}$ and $10^{-15}$~G, depending on the adopted model source (from ``minimal'' to ``maximal'', through all the ``intermediate'' possibilities) and the model of the EBL. The tighest bound is derived from the data on 1ES~0229+200, at the level of $10^{-15}$~G, which is consistent with the bounds found under similar assumptions about the cascade suppression mechanism by \citet{Neronov:2009,Dolag:2010ni,Tavecchio:2010ja}. We stress that the bound should be considered only as an order-of-magnitude estimate, due to the significant uncertainty of the shape and overall normalization of the cascade introduced by the uncertainty of the normalization and spectral shape of the EBL.

For the case of suppression of cascade emission due to the time delay of the cascade signal, one assumes that the primary source is active only during a limited period of time, just about the time span of gamma-ray observations ($t_{\rm source}\sim 1$~yr). Time delay of the cascade signal by $t_{\rm delay}>t_{\rm source}$ would lead to the suppression of the cascade flux by a factor $t_{\rm source}/t_{\rm delay}$. Figs. \ref{suppression_RGB}, \ref{suppression_1ES_0229}, and \ref{suppression_1ES_1218} show  that time delay of the cascade signal starts to influence the cascade emission signal at GeV energies when the magnetic field strength reaches $\sim 10^{-18}$~G. Similar to the case of suppression due to the extended emission, the precise value of $B$ necessary to suppress the cascade emission depends on the adopted source and EBL models.  The tighest lower bound is again derived from the data on 1ES~0229+200, at the level of $10^{-17}$~G. This bound should also be considered as an order-of-magnitude estimate because of the remaining uncertainty in the measurements of the spectrum of EBL.

Note that the bound $B\ge 10^{-17}$~G derived assuming suppression of cascade emission due to the time delay of the cascade signal in the case of 1ES~0229+200 is by 1.5 orders of magnitude stronger than the bound derived from a similar analysis of the same source by \citet{Dermer:2010mm}. We believe that the main source of discrepancy between the result obtained in the present work and that of \citet{Dermer:2010mm} is the simplified analytical treatment of the cascade emission adopted by  \citet{Dermer:2010mm}. The simplified treatment of the cascade has led to an under-estimate of the cascade flux at high energies $E_{\gamma}\gtrsim 10$~GeV and an over-estimate of the strength of suppression of the cascade emission due to the time delay at low energies $E_{\gamma}\lesssim 10$~GeV. 


Furthermore, we note that our limit of $B>10^{-17}$~G from the time delay of the cascade signal is consistent with the results of a similar analysis  by \citet{Dolag:2010ni}, who found somewhat tighter bound $B>10^{-16}$~G, assuming a larger minimal possible time delay, $t_{\rm delay}>100$~yr in the cascade emission from 1ES~0229+200. 

A summary of the limits on magnetic fields in the intergalactic medium, which can be derived from the simultaneous GeV-TeV band observations is shown in Fig. \ref{exclusion_time}. In our analysis we have considered the bound on the EGMF strength assuming a fixed magnetic field correlation length $\lambda_B=1$~Mpc. If the EGMF correlation length is $\lambda_B\gtrsim 1$~Mpc, the lower bound on EGMF strength does not depend on $\lambda_B$ because the cooling distance of $e^+e^-$ pairs is much shorter than the typical size of regions in which EGMF is correlated. We have explicitly verified  this by making a control run of Monte-Carlo simulations with $\lambda_B=30$~Mpc and comparing the results with the case $\lambda_B=1$~Mpc shown above. On the other hand, if $\lambda_B\lesssim 1$~Mpc, the inverse Compton cooling distance becomes larger than the size of the regions with correlated EGMF. This means that electrons and positrons pass through  regions with different magnetic field orientations during their cooling. As a result, the deflection angle scales proportionally to the square root, rather than linearly with the propagation distance on the distance scales comparable to the inverse Compton cooling length. This explains the improvement of the lower bound on the EGMF strength $B\sim \lambda_B^{-1/2}$ at $\lambda_B\lesssim 1$~Mpc: stronger magnetic field is required to deviate electron trajectories by a given angle.

\section{Conclusion}
\label{Conclusion}

In this paper we have derived constraints on the strength of magnetic fields in the intergalactic medium from simultaneous observations of blazars in the GeV band (by {\it Fermi}/LAT telescope) and TeV band (by ground-based $\gamma$-ray telescopes). The constraints stem from the requirement that the GeV band signal from electromagnetic cascade initiated by the absorption of the primary TeV $\gamma$-rays in interactions with Extragalactic Background Light should be suppressed by deflections of electron-positron pairs by magnetic fields in the intergalactic medium. Non-observation of the cascade emission by {\it Fermi}/LAT telescopes imposes a lower bound on the cascade flux suppression factor which could be converted to a correlation length dependent lower bound on the strength of magnetic field.

We have found that constraints on the magnetic field strength could be derived from the $\gamma$-ray data on three blazars, 1ES~0229+200, RGB~J0710+591 and 1ES~1218+304 (out of seven, for which simultaneous GeV-TeV data are available). For all three sources, we have performed detailed modeling of the spectral characteristics in the broad (0.1 GeV to 10 TeV) energy range. We have fitted the observed $\gamma$-ray spectra with a two-component spectral model which consists of both direct absorbed emission from the blazar and a cascade emission component calculated using detailed Monte-Carlo simulations of the cascade development. The observational data are inconsistent with the models in which the cascade emission is calculated assuming zero magnetic field strength in the cascade development region (extending to $\sim 100$~Mpc distance from the source along the line of sight). 

The minimal magnetic field strength required to achieve sufficient suppression of the cascade signal depends on the assumption about the mechanism of suppression of the cascade signal. If the suppression is due to the time delay of the cascade emission following a period of enhanced activity of the source in the TeV band (with duration $\sim 1$~yr), then the minimal required field strength is $B\sim 10^{-17}$~G in the case of the field with large correlation length $\lambda_B\gtrsim 1$~Mpc. If the (unknown)  correlation length is   $\lambda_B\lesssim 1$~Mpc, the minimal needed magnetic field strength is larger by a factor $\left(\lambda_B/1\mbox{ Mpc}\right)^{-1/2}$. If the suppression of the cascade emission is due to the fact that the size of the cascade source is much larger than the point spread function of {\it Fermi}/LAT telescope, rather than due to the time delay of the cascade signal, the lower bound is $B\ge 10^{-15}$~G, with the same dependence of the correlation length $\lambda_B$. 

The two possibilities for suppression of the cascade emission could be distinguished via a search of the delayed GeV $\gamma$-ray emission following strong TeV band flares of blazars or of the extended emission around TeV blazars in the GeV-TeV band. The distinguishing feature of the delayed cascade emission is the characteristic energy dependence $t_{\rm delay}\sim E_\gamma^{-2.5}$ or $\sim E_{\gamma}^{-2}$, as shown in Fig. \ref{t_delay}. If the real magnetic field strength is close to $B\sim 10^{-15}$~G, the extended emission around TeV blazars should be detectable with {\it Fermi}, while the time delay of the cascade emission might be detectable at higher energies ($\sim 100$~GeV) by the ground-based $\gamma$-ray telescopes.

We have investigated the dependence of the derived limit on the assumptions about the intrinsic spectrum of the sources by considering the extreme cases of ``minimal'' and ``maximal'' cascade contributions. In the ``minimal'' case the parameters of the intrinsic spectrum of the source are chosen in such a way that they minimize the total flux in the cascade component (at zero magnetic field strength). In the ``maximal'' cascade model the cascade flux dominates over the intrinsic source flux at all energies up to the TeV range. Surprisingly, the lower bound on the magnetic field strength is practically independent on the choice of the model used to fit the observed GeV -- TeV band spectra. This is explained by the fact that the amount of power transferred by the cascade from the TeV to the GeV energy band is determined only by the measured TeV flux from the direction of the source and is not sensitive to the origin of the TeV $\gamma$-rays (if they are intrinsic to the source or produced  in the course of development of electromagnetic cascade close to the source). Since ``minimal'' and ``maximal'' cascade models represent the two extreme possibilities for the possible amount of cascade contribution to the source flux, we conclude that the uncertainty of the modeling of the observed source spectrum introduces an uncertainty by a factor of $\sim 1$ in the derived lower bound on EGMF strength.

\bibliographystyle{aa} 

\end{document}